%% file: ms.tex
\documentclass[iop]{emulateapj}

\usepackage{natbib}
\usepackage{amsmath}
\usepackage{url}
\usepackage{graphicx}

\begin{document}

\title{The WISE AGN Catalog}

\author{R.J.~Assef}
\affiliation{N\'ucleo de Astronom\'ia de la Facultad de Ingenier\'ia,
  Universidad Diego Portales, Av. Ej\'ercito Libertador 441, Santiago,
  Chile. Email: roberto.assef@mail.udp.cl}

\author{D.~Stern}
\affiliation{Jet Propulsion Laboratory, California Institute of
  Technology, 4800 Oak Grove Drive, Pasadena, CA 91109, USA.}

\author{G.~Noirot}
\affiliation{Jet Propulsion Laboratory, California Institute of
  Technology, 4800 Oak Grove Drive, Pasadena, CA 91109, USA.}
\affiliation{Universit\'e Paris-Diderot Paris VII, Universit\'e de
  Paris Sorbonne Cit\'e ( PSC ), F-75205 Paris Cedex, France.}

\author{H.D.~Jun}
\affiliation{Jet Propulsion Laboratory, California Institute of
  Technology, 4800 Oak Grove Drive, Pasadena, CA 91109, USA.}
\affiliation{School of Physics, Korea Institute for Advanced Study, 85
  Hoegiro, Dongdaemun-gu, Seoul 02455, Korea}

\author{R.M.~Cutri}
\affiliation{IPAC, Mail Code 100-22, California Institute of
  Technology, 1200 E. California Blvd, Pasadena, CA 91125, USA.}

\author{P.R.M.~Eisenhardt}
\affiliation{Jet Propulsion Laboratory, California Institute of
  Technology, 4800 Oak Grove Drive, Pasadena, CA 91109, USA.}

\begin{abstract}
We present two large catalogs of AGN candidates identified across
30,093 deg$^2$ of extragalactic sky from the Wide-field Infrared
Survey Explorer's AllWISE Data Release.  Both catalogs are selected
purely using the WISE W1 and W2 bands. The R90 catalog consists of
4,543,530 AGN candidates with 90\% reliability, while the C75 catalog
consists of 20,907,127 AGN candidates with 75\% completeness. These
reliability and completeness figures were determined from a detailed
analysis of UV- to near-IR spectral energy distributions of $\sim
10^5$ sources in the 9~deg$^2$ Bo\"otes field.  The AGN selection
criteria are based on those of \citet{assef13}, re-calibrated to the
AllWISE data release.  We provide a detailed discussion of potential
artifacts, and excise portions of the sky close to the Galactic
Center, Galactic Plane, nearby galaxies, and other expected
contaminating sources. These catalogs are expected to enable a broad
range of science, and we present a few illustrative cases.  From the
R90 sample we identify 45 highly variable AGN lacking radio
counterparts in the FIRST survey.  One of these sources,
WISEA~J142846.71+172353.1, is a changing-look quasar at $z=0.104$,
which has changed from having broad H$\alpha$ to being a narrow-lined
AGN. We characterize our catalogs by comparing them to large,
wide-area AGN catalogs in the literature. We identify four {\it ROSAT}
X-ray sources that each are matched to three WISE-selected AGN in the
R90 sample within 30\arcsec. Spectroscopy reveals one of these
systems, 2RXS~J150158.6+691029, consists of a triplet of quasars at
$z=1.133\pm0.004$, suggestive of a rich group or forming galaxy
cluster.
\end{abstract}


\section{Introduction}

Most UV through near-IR emission constituting the Spectral Energy
Distribution (SED) of an AGN is produced by the innermost regions of
the accretion disk, which spans distances down to the last stable
orbital radius of the black hole \citep[e.g.,][]{shakura73}. At larger
distances from the accretion disk, a dusty medium, usually referred to
as the ``dust torus'' \citep[see,
  e.g.,][]{antonucci93,urry95,netzer15}, absorbs the light of the
accretion disk and re-emits it in the infrared, dominating the SED at
wavelengths longer than $\sim$1~$\mu$m. Initially assumed to be a
smooth dust structure with a toroidal geometry (hence its name),
observations suggest that the dust is more likely found in
geometrically and optically thick clouds, toroidally distributed
around the central engine
\citep[e.g.,][]{krolik88,nenkova02,nenkova08, elitzur06,tristram07},
although a number of uncertainties remain about the exact properties
and distribution of the dust \citep[e.g.,][]{feltre12,netzer15}. The
dust is heated by the accretion disk emission, with the inner boundary
of the torus being set by the sublimation temperature of the
dust. Given the high temperatures of the torus, its emission is most
prominent in the shorter mid-IR wavelengths ($\lesssim$ 50~$\mu$m). At
longer wavelengths, the observed emission can become dominated by the
cold dust of the host galaxy, typically associated with
star-formation. Because of the torus and accretion disk emission, the
mid-IR is ideal for AGN identification, as its SED is very different
than that of stars and inactive galaxies. At low redshifts, the SED in
the observed mid-IR bands is dominated by emission from the dust
torus, while at higher redshifts the mid-IR bands map the
optical/near-IR accretion disk emission.

Along with mid-IR selection, the other most successful methods of AGN
identification are arguably those based on X-ray observations, and
those based on UV and optical broad-band photometry and spectra. Each
of these wavelengths has different advantages and disadvantages, and
obtains samples with different biases. For example, optical AGN
identification is severely affected by dust obscuration (either from
the torus or the host galaxy), making samples heavily biased against
type 2 (or obscured) AGN, while both mid-IR and X-ray identification
are much more robust against obscuration, particularly higher energy,
or hard X-ray identification. Furthermore, mid-IR and optical
identification can be diluted significantly by emission from the host
galaxy, as host light can be very significant at these wavelengths,
rivaling the AGN emission in many cases. As host emission light is
related to the black hole mass \citep[e.g.,][]{marconi03}, this
translates into samples that are biased against AGN accreting at small
fractions of their Eddington limit
\citep[e.g.,][]{hickox09,assef11,mendez13}. On the other hand,
optical/UV broad-band photometry can be efficiently obtained by
ground-based telescopes, making the observations much easier than in
the X-rays and the mid-IR. While mid-IR observations can be obtained
from the ground for the brightest targets in the sky, broad-band
photometry to identify large AGN samples can only be efficiently
obtained by space-based observatories due to the Earth's
atmosphere. X-ray observations can only be obtained by space-based
facilities, but they require significantly longer exposure time than
space-based mid-IR observations. For example, \citet{gorjian08} finds
that 97.5\% of X-ray sources identified in the 5~ks XBo\"otes survey
{\it{Chandra}} observations \citep{murray05} have counterparts in the
90~s observations of the {\it{Spitzer}} IRAC Shallow Survey
\citep{eisenhardt04}, with an additional $\sim$1\% of the X-ray
sources expected to be spurious. Despite some of its shortcomings,
mid-IR AGN identification is very important for a thorough census of
AGN activity, being significantly less biased than UV/optical
identification, while requiring significantly less observing time than
X-ray identification.

For all of these reasons, the Wide-field Infrared Survey Explorer
\citep[WISE;][]{wright10} is an ideal mission to identify a very large
number of AGN across the full sky. With its 40~cm aperture, the WISE
mission imaged the entire sky in four mid-IR bands, centered at 3.4,
4.6, 12 and 22~$\mu$m, referred to as W1, W2, W3 and W4,
respectively. The FWHM of the point spread function (PSF) in the W1,
W2 and W3 bands is 6\arcsec\ while in the W4 band it is
12\arcsec. WISE is in a polar orbit, requiring approximately 6 months
to scan the entire sky. The cryogenic survey was conducted between
January and August 2010, and completed slightly more than one pass
over the entire sky. After the exhaustion of cryogen, NASA's Planetary
Division funded an extension to focus on near-Earth objects
\citep{mainzer11}, which continued observing in the W1 and W2 bands
until completing a second pass over the entire sky in February
2011. All data obtained by WISE from 2010 and 2011 has been made
public in the AllWISE Data
Release\footnote{\url{http://wise2.ipac.caltech.edu/docs/release/allwise/}},
and we use this data set as the starting point to construct the mid-IR
AGN catalogs presented here.

A substantial number of mid-IR AGN identification techniques have been
developed in the literature. While the initial techniques were
developed already for the IRAS satellite observations
\citep[e.g.,][]{degrijp85,degrijp87,leech89}, which provided the first
infrared survey of the sky, the majority have been developed for the
more recent observatories, such as {\it{Spitzer}} and WISE
\citep{lacy04,lacy07,lacy13,stern05,stern12,assef10,assef13,jarrett11,donley12,mateos12,messias12,wu12b}. \citet{stern12}
studied the WISE colors of AGN in the 2~deg$^2$ Cosmic Evolution
Survey \citep[COSMOS;][]{scoville07} field, relying on its earlier
deep {\it{Spitzer}}/IRAC observations for the AGN
identification. \citet{stern12} were able to define WISE AGN
selection criteria based solely on the W1 and W2 magnitudes, showing
that down to a W2 magnitude of $15.05$ (10$\sigma$ detection at the
ecliptic latitude of the COSMOS field), 78\% of
{\it{Spitzer}}-identified AGN have W1$-$W2$>0.8$, and that 95\% of the
objects with such red WISE colors are {\it{bona fide}} AGN. Using this
criterion, \citet{stern12} identified 61.9$\pm$5.4 AGN per
deg$^2$. Motivated by these results, \citet[][A13 hereafter]{assef13}
expanded such studies to the larger 9~deg$^2$ NOAO Deep, Wide-Field
Survey \citep[NDWFS;][]{jannuzi99} Bo\"otes field. Because its higher
ecliptic latitude as compared to the COSMOS field has denser coverage
from the WISE survey (see \S\ref{sssec:smooth_gradient}), the NDWFS
Bo\"otes field also allowed us to probe the AGN selection to
significantly deeper WISE magnitudes. Using the extensive
UV-through-mid-IR photometric and spectroscopic observations available
for this field (see A13 for detailed account of the data), we were
able to reliably identify AGN down to W2$=$17.11 (3$\sigma$
detection), thereby extending and improving the WISE AGN selection,
and providing different selection criteria separately optimized for
reliability and completeness. In particular, the criteria optimized
for 90\% reliability, referred to as R90, yields a surface density of
130$\pm$4 AGN candidates per deg$^2$.

Relying on different selection criteria, AGN catalogs based on WISE
observations have been published by several other authors
\citep[e.g.,][]{edelson12,secrest15,dipompeo15}. In this work we apply
the selection method devised by A13 to generate the largest AGN
catalog based on the AllWISE data release. In
\S\ref{sec:agn_selection} we re-calibrate the selection function of
A13 to the AllWISE data using the same set of multi-wavelength
observations in the NDWFS Bo\"otes field, as there are significant
improvements in the photometry from the All-Sky to the AllWISE data
releases. In \S\ref{sec:agn_cats} we discuss the generation of two
WISE AGN catalogs, respectively based on reliability- and
completeness-optimized selections. We also include a discussion of the
spatial filters applied, and discuss the general properties of these
catalogs. In \S\ref{sec:agn_var} we discuss the highest variability
sources in the reliability-optimized catalog, and in
\S\ref{sec:other_cats} we compare our AGN catalogs with large AGN
catalogs in the literature. We assume a flat $\Lambda$CDM cosmology
with $\Omega_M=0.3$, $\Omega_{\Lambda}=0.7$ and $H_0=70~\rm km~\rm
s^{-1}~\rm Mpc^{-1}$. All photometry is presented in the natural
photometric system of their bands unless stated otherwise (i.e., AB
for $griz$ and Vega for the rest, i.e., $B_w$, $R$, $I$, $J$, $H$,
$Ks$, $K$ as well as the {\it{Spitzer}} and the WISE bands).

\section{The AGN Selection Criteria}\label{sec:agn_selection}

The selection criteria we use to produce the WISE AGN catalogs
presented in \S\ref{sec:agn_cats} is based upon selection criteria
developed by A13 using AGN in the NDWFS Bo\"otes field. A13 presented
four distinct AGN selection criteria based only on the W1 and W2
magnitudes of the sources, chosen based on the results of
\citet{stern12}, not requiring detections in the lower sensitivity W3
and W4 bands. Two of the criteria presented by A13 were aimed at
producing catalogs with 90\% and 75\% reliability (referred to as the
R90 and R75 criteria, respectively), while the other two were aimed at
yielding 90\% and 75\% completeness (C90 and C75,
respectively). Specifically, the two reliability optimized AGN
    selection criteria of A13 are given by:
\begin{equation}
  \rm W1 - \rm W2 > \alpha_{\rm R} \exp \{\beta_{\rm R}(\rm
    W2-\gamma_{\rm R})^2\},
\end{equation}
with $(\alpha_{R90},\beta_{R90},\gamma_{R90})=(0.662, 0.232, 13.97)$
and $(\alpha_{R75},\beta_{R75},\gamma_{R75})=(0.530, 0.183,
13.76)$. The two completeness optimized AGN selection criteria of A13
are in turn given by:
\begin{equation}
  \rm W1 - \rm W2 > \delta_{\rm C},
\end{equation}
with $\delta_{\rm C90}=0.50$ and $\delta_{\rm C75}=0.77$. A13 observed
that in order to obtain highly reliable samples at increasingly
fainter W2 magnitudes, redder W1--W2 colors were required due to a
combination of evolution in the contamination by non-active galaxies
and the larger uncertainties at fainter magnitudes. Conversely, the
completeness fractions for a given W1--W2 color cut appeared to be
independent of magnitude. Hence, the functional forms of the R90 and
R75 criteria have strong dependencies on the W2 magnitudes while the
C90 and C75 criteria solely rely on W1--W2 color boundaries. In fact,
the C75 criterion is nearly identical to that proposed by
\citet{stern12} for brighter magnitudes, namely, $\rm W1 - \rm W2 >
0.80$.

In the following sections we will present two AGN catalogs, one
optimized for reliability and one for completeness. These catalogs are
respectively based on modified versions of the R90 and C75 criteria of
A13. Modifications are needed over the criteria presented by A13
because they used the earlier WISE All-Sky data
release\footnote{\url{http://wise2.ipac.caltech.edu/docs/release/allsky/}}
for their study, while in this study we use the newer AllWISE data
release. The All-Sky data release is limited to the data obtained
during the cryogenic mission, while the AllWISE release incorporates
data obtained during the post-cryogenic main mission extension, known
as NEOWISE. Furthermore, \citet{lake13} shows that WISE All-Sky
profile fitting fluxes of faint sources are underestimated by
7$\pm$2~$\mu$Jy and 11$\pm$2~$\mu$Jy in W1 and W2, respectively, due
to excessive sky subtraction, an issue that has been corrected in the
AllWISE data release.

To re-calibrate the WISE AGN selection criteria developed by A13 we
use the same auxiliary photometric data sets available in the NDWFS
Bo\"otes field and follow the same analysis steps. We refer the reader
to A13 and \citet{assef10} for a detailed account of the auxiliary
photometric and spectroscopic data sets used and of the methods used
to derive the SED classifications and photometric redshifts when no
spectroscopic ones were available. In summary, the photometric
broad-band data spans the UV to the mid-IR with very good sampling. In
addition to the original deep $B_w$, $R$, $I$ and $K$ broad-band
imaging from the NDWFS survey, we also use data from the NUV and FUV
bands of {\it{GALEX}} \citep{martin05}, $z$-band from the zBo\"otes
survey \citep{cool07}, $J$, $H$ and $Ks$ bands of NEWFIRM
\citep{gonzalez10}, the {\it{Spitzer}}/IRAC [3.6], [4.5], [5.8] and
      [8.0] bands from SDWFS \citep{ashby09}, and {\it{Spitzer}}/MIPS
      24$\mu$m data from MAGES \citep{jannuzi10}. Specifically, we use
      6\arcsec\ diameter aperture magnitudes, corrected for PSF losses
      and obtained from PSF-matched images in all but the {\it{GALEX}}
      and {\it{Spitzer}} bands. With the exception of the {\it{GALEX}}
      and MIPS data, source photometry is extracted from all images at
      the positions of [4.5] sources. Photometry from those two
      catalogs were obtained from positional matching. The
      spectroscopic redshifts come mainly from the AGN and Galaxy
      Evolution Survey \citep[AGES;][]{kochanek12}, which obtained
      deep optical spectra of $23,745$ sources in the field, and are
      supplemented with deeper spectroscopy of $\sim 2,000$ sources
      obtained with various facilities, although primarily from Keck
      \citep[e.g.,][]{eisenhardt08}.

We start with the sources listed in the AllWISE catalog of objects in
the NDWFS Bo\"otes field, obtained through the NASA/IPAC Infrared
Science Archive
(IRSA\footnote{\url{http://irsa.ipac.caltech.edu/}}). The AllWISE
Source Catalog consists of sources detected with $\rm SNR>5$ in at
least one band and not flagged as spurious detections, among other
criteria, and provides all magnitudes with a significance of at least
2$\sigma$. We direct the reader to the AllWISE
documentation\footnote{\url{http://wise2.ipac.caltech.edu/docs/release/allwise}}
for details. In addition, for this experiment we further require all
of our sources to: (i) be detected at the $3\sigma$ level in W1 and at
the $5\sigma$ level in W2; (ii) be point sources ({\tt{ext\_flg=0}});
(iii) not be contaminated by image artifacts in any band
({\tt{cc\_flags=0000}}); and (iv) not be blended with other sources
({\it{nb=1}}). For the WISE AGN catalog presented in the following
sections, we relax requirements (i), (iii) and (iv), but we enforce
them here when defining the selection criteria.

We cross-match the positions of the WISE sources with sources in the
auxiliary photometric catalogs described above using a
2\arcsec\ matching radius. While somewhat conservative when
considering the width of the WISE PSF, we adopt this matching radius
as A13 found it to work well for matching WISE data to the data sets
described earlier. Approximately 4\% of WISE sources do not have
matches in the {\it{Spitzer}} [4.5] catalog, usually because of source
blending in the lower resolution WISE images \citep{stern12}. Using
the full, broadband multi-wavelength data, we determine which sources
are AGN based on their SEDs. As discussed in A13, we use the SED
fitting algorithm and templates of \citet{assef10} to obtain
photometric redshifts for sources lacking spectroscopic redshifts, and
model the SEDs of all sources in the field. Specifically, each source
is modeled as a non-negative linear combination of three galaxy SED
templates, resembling respectively E, Sbc and Im galaxies, and an AGN
SED template while also fitting for its redshift. We also fit for the
reddening of the AGN SED template with a weak prior that punishes
large obscurations, and we include IGM absorption for all four
templates using the prescription outlined in \citet{assef10}. These
templates span the wavelength range of 0.03--30$\mu$m and were
iteratively derived by \citet{assef10} from the UV (rest-frame
0.03\AA) through mid-IR photometry of 14,448 galaxies and 5,347 likely
AGN with spectroscopic redshifts from the AGES survey in this
field. To derive the photometric redshifts we also apply a luminosity
prior based on the Las Campanas Redshift Survey \citep{lin96}
luminosity function that only affects the galaxy templates. We
conservatively consider as AGN all sources with $\hat{a}>0.5$, where
$\hat{a}$ is defined as the fraction of the 0.1--30$\mu$m luminosity
coming from the AGN component, after correcting the latter for
obscuration, namely:
\begin{equation}
  \hat{a} = \frac{L_{\rm AGN}}{L_{\rm AGN} + L_{\rm Host}}.
\end{equation}
\noindent We refer the reader to \citet{assef10}, \citet{chung14} and
A13 for details on the SED modeling and the analysis. Note that the
$\hat{a}$ parameter is relatively robust to uncertainties in the
redshift, which is particularly important given the large
uncertainties of AGN photometric redshifts \citep{assef10}.

\begin{figure}
  \begin{center}
    \plotone{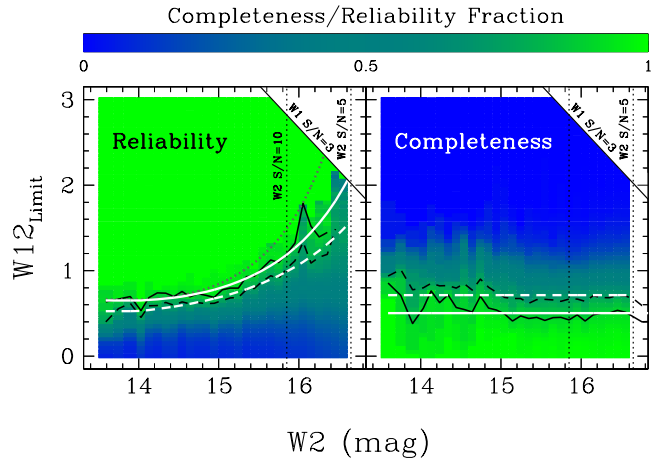}
    \caption{Reliability ({\it{left panel}}) and completeness
      ({\it{right panel}}) of AGN candidates defined by $\hat{a}>0.5$
      selected by the color cut $\rm W1-\rm W2>\rm W12_{\rm Limit}$ as
      a function of $\rm W2$ magnitude. Reliability and completeness
      of 90\% (75\%) are shown as a function of magnitude by the solid
      (dashed) black lines. Objects redder than the top right corner
      of the panels are missing due to the W1 $S/N > 3$
      requirement. The proposed reliability optimized criteria
      (eqn.[\ref{eq:rel_sel}]) for 90\% (R90) and 75\% (R75)
      reliability are shown in the left panel by the white solid and
      dashed lines, respectively. For comparison, the dotted gray line
      shows the R90 criterion of A13. The completeness-optimized
      criteria (eqn.[\ref{eq:comp_sel}]) for 90\% (C90) and 75\%
      (C75) completeness are shown in the right panel with the
      same respective line styles as in the left panel.}
    \label{fg:agn_sel}
  \end{center}
\end{figure}

Our primary aim here is to re-calibrate the mid-IR AGN selection of
A13 based on the W1$-$W2 color and W2 magnitude. In general, AGN are
easily identified in the W1 and W2 bands because they are
significantly redder than galaxies at the depth of the WISE survey in
the NDWFS Bo\"otes field, so the main criterion to select AGN can be
written as $\rm{W1}-\rm{W2} > \rm{W12}_{\rm Limit}$. Indeed,
\citet{stern12} showed in the COSMOS field that for W2$<$15.05,
$\rm{W12}_{\rm Limit}=0.8$ yields an AGN sample that is 95\% reliable
and 75\% complete, while A13 showed that the contamination rate for a
given $\rm{W12}_{\rm Limit}$ is a strong function of W2 at fainter
magnitudes. With this in mind, Figure \ref{fg:agn_sel} shows the
reliability (left panel) and completeness (right panel) of AGN samples
selected for a given $\rm{W12}_{\rm Limit}$ as a function of W2
magnitude. The general behavior is consistent with that described by
A13, implying that to maintain the sample reliability it is necessary
to adopt a functional form of $\rm{W12}_{\rm Limit}$ that depends on
W2, while a fixed value of $\rm{W12}_{\rm Limit}$ is appropriate for
maintaining a given AGN sample completeness. Hence, it is necessary to
develop different selection criteria depending on whether the primary
goal is to optimize the reliability or the completeness of the
WISE-selected AGN sample.

We model the AGN selection criteria using the same functional forms of
A13, with one minor modification. For the reliability-optimized
criteria, we consider the following form
\begin{equation}\label{eq:rel_sel}
  \rm{W1}-\rm{W2}\ >\ \left\{
  \begin{array}{rl}
    \alpha_R\ \exp\{\beta_R(\rm W2 - \gamma_{\rm R})^2\}, &
    \rm{W2}>\gamma_R\\ 
    \alpha_R, & \rm{W2}\leq \gamma_R
  \end{array}\right. ,
\end{equation}
\noindent where the value of the $\alpha_R$, $\beta_R$ and $\gamma_R$
depend on the reliability fraction targeted.  This form of the
selection criteria is equivalent to that used by A13 for W2 magnitudes
fainter than $\gamma_R$, while it stops evolving with W2 for brighter
magnitudes. A13 neglected to specify the constant term for bright
magnitudes, simply presenting the term for fainter magnitudes. The
number of point sources at magnitudes bright enough where this is an
issue is very small, and certainly has no effect on the results
presented by A13. However, since the goal of this work is to present
an AGN sample across most of the sky, we correct this detail. Using
the results of Figure \ref{fg:agn_sel}, we find that a reliability of
90\% is achieved by
$(\alpha_{R90},\beta_{R90},\gamma_{R90})=(0.650,0.153,13.86)$, while a
reliability of 75\% is achieved by
$(\alpha_{R75},\beta_{R75},\gamma_{R75})=(0.486,0.092,13.07)$. Both
criteria are shown in the left panel of Figure \ref{fg:agn_sel}. For
comparison, the figure also shows the R90 criterion of A13
obtained using the WISE All-Sky Data Release instead of the AllWISE
Data Release. The much steeper dependence on W2 magnitude of the A13
criteria is expected due to the previously mentioned flux bias present
in the All-Sky Data Release \citep{lake13}. We find that the R90 and
R75 criteria have a completeness of 17\% and 28\% respectively.

For the completeness-optimized criteria we use the functional form of
\citet{stern12} and A13, namely
\begin{equation}\label{eq:comp_sel}
  \rm W1 - \rm W2 > \delta_C,
\end{equation}
\noindent and we find that 75\% completeness is achieved for
$\delta_{C75}=0.71$ while 90\% completeness is achieved for
$\delta_{C90}=0.50$. Both criteria are shown in the right panel of
Figure \ref{fg:agn_sel}. While $\delta_{C90}$ has the same value as
found by A13, $\delta_{C75}$ is 0.06~mag bluer than that found by
A13. This is also a likely consequence of the flux bias in the All-Sky
Data Release, as the value of $\rm W12_{\rm Limit}$ seems
systematically bluer for $\rm W2\gtrsim 15~\rm mag$. We find that the
C90 and C75 criteria have a reliability of 34\% and 51\%
respectively. In the next section we present WISE-selected AGN
catalogs across $\sim$75\% the sky based on the R90 and C75
criteria derived here.

\begin{figure}
  \begin{center}
    \plotone{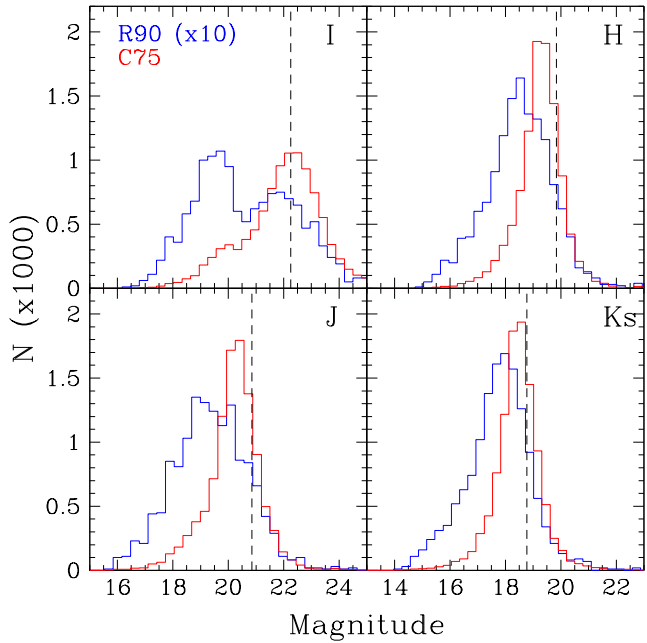}
    \caption{Magnitude distribution of the R90 (blue) and C75 (red)
      selected AGN in the NDWFS Bo\"otes field in the $I$ (top-left
      panel), $J$ (bottom-left panel), $H$ (top-right panel) and $Ks$
      (bottom-right panel) bands. The vertical dashed lines show the
      approximate magnitude at which the $S/N$ in the given band is
      equal to 3.}
    \label{fg:bootes_mags}
  \end{center}
\end{figure}

\begin{figure}
  \begin{center}
    \plotone{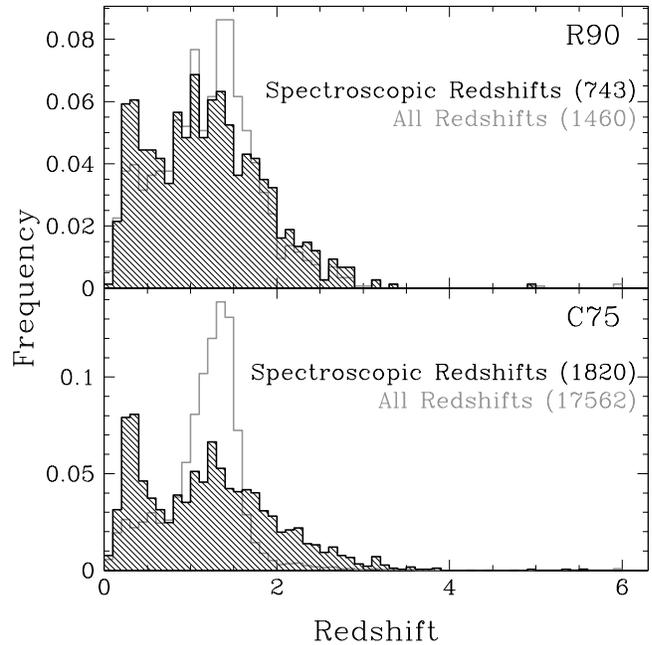}
    \caption{Redshift distribution of the sources selected by the R90
      (top) and C75 (bottom) criteria in the NDWFS Bo\"otes field. The
      black hashed histograms show the distribution of spectroscopic
      redshifts and the gray histograms show the distribution of
      spectroscopic and photometric redshifts combined.}
    \label{fg:bootes_z}
  \end{center}
\end{figure}

Figure \ref{fg:bootes_mags} shows the magnitude distribution in the
$I$, $J$, $H$ and $Ks$ bands of the R90 and C75 selected AGN in the
Bo\"otes field. As expected, the R90 sample is brighter on average
than the C75 sample in all four bands. Interestingly, the $I$-band
distribution is bimodal, reflecting the fact that our criteria selects
unobscured as well as obscured AGN. The near-IR bands, on the other
hand, do not show this behavior, consistent with the fact that redder
bands are less affected by obscuration. Figure \ref{fg:bootes_z} shows
the redshift distribution for the R90 and C75 selected AGN in the
NDWFS Bo\"otes field, primarily obtained by AGES
\citep{kochanek12}. The bimodality of the distribution is likely
caused by the fact that at lower redshifts mid-IR selection is more
sensitive to obscured AGN than at higher redshifts. The prominent
photometric redshift peak at $1\lesssim z \lesssim 2$ observed for the
C75 sample is likely due to the contribution of contaminating
elliptical galaxies as well as of real AGN too faint for spectroscopic
redshifts. As expected, the number of sources in both catalogs
declines for $z\gtrsim 2$ and very few are found at $z\gtrsim 3$. This
is likely caused in part by the characteristics of the spectroscopic
follow-up as well as by the WISE colors becoming progressively bluer
with redshift in the range $2\lesssim z\lesssim 5$ (see, e.g., Fig. 1
of A13). We refer the reader to A13 (and references therein) for a
discussion of the spectroscopic sample and the photometric redshift
reliability.

\section{The WISE AGN Catalogs}\label{sec:agn_cats}

Using the selection criteria presented in \S\ref{sec:agn_selection} we
construct a map of WISE-selected AGN across $\sim$75\% of the sky. We
construct two different catalogs, a reliability-optimized one based on
the R90 selection criterion, and a completeness-optimized one based on
the C75 criterion. We only consider sources with W1 and W2 magnitudes
fainter than the saturation limits of the survey (i.e., W1$>$8 and
W2$>$7) and with $S/N>5$ in W2, classified as point sources and not
flagged as either artifacts or affected by artifacts (i.e., we require
that the {\tt{cc\_flags}} parameter is 0 in both W1 and W2). We refer
to these as the ``raw catalogs'', since these catalogs are affected by
a number of contaminants not present in the NDWFS Bo\"otes field due
to its size and high Galactic latitude. Next, we discuss several
spatial filters applied to the raw catalogs designed to limit the
number of such contaminants in the final catalogs.

\subsection{Spatial Filters}\label{ssec:spatial_filters}

\subsubsection{The Galactic Plane and the Galactic Center}

The NDWFS Bo\"otes field is an extragalactic field, centered
approximately 67 deg away from the Galactic Plane (GP) and 77
deg away from the Galactic Center (GC). This limits the number of
stellar contaminants that could affect our sample in the Bo\"otes
field, such as young stellar objects (YSOs), asymptotic giant branch
(AGB) stars, and H{\,\sc ii} regions, as well as possible artifacts
arising from the high concentration of sources in regions near the GP
and the GC.

To avoid these issues, the first spatial filter we apply removes all
sources closer than 30 deg from the GC, and all sources closer
than 10 deg from the GP. We chose these cuts following the
approach of \citet{eisenhardt12}, who used them to select Hot
Dust-Obscured Galaxies in the WISE data, although the general
properties of the final catalog should be insensitive to small changes
in these parameters. The area removed by these cuts is 8,753 deg$^2$.

\citet{nikutta14} finds that YSOs are primarily distributed within 6
degrees of the GP, and hence the above cut should eliminate the
majority of these sources. The remainder should be associated with
star-forming regions at higher Galactic latitudes, which we discuss
further in \S\ref{sssec:hii_and_sf_regions}. \citet{nikutta14} also
studied the colors of AGB stars in the WISE bands, and panel 4 of
their Figure 8 shows that the majority of these sources have
W1--W2$\lesssim$0.5, implying that only a small fraction would make it
into our AGN catalogs. While \citet{jackson02} estimates that our
Galaxy contains approximately 200,000 AGB stars in total, most of them
should be close to the GP and GC. Specifically, their Figure 8 shows
that the great majority of the sources in their sample are within 10
degrees of the GP, implying that the above cut should eliminate the
majority of these sources from our AGN catalogs.

\citet{secrest15} reports effects of the GP up to 15 deg in Galactic
latitude. To test this, we studied the surface density of sources with
W2$<$15 as a function of distance to the GP in several slices of
Galactic longitude. We apply this magnitude limit as AllWISE achieves
this depth in all regions in the sky farther than 10 deg away from the
GP. We find that the main issue caused by the GP is a noticeably lower
surface density due to source confusion noise. This effect can be
observed up to $\sim 40$~deg away from the GP and is progressively
more severe closer to the GP. At a distance of $\sim$15~deg from the
GP, the source density is about 50\% of that in the high Galactic
latitude sky. This implies that the completeness of our catalog is
lower for low Galactic latitudes, although reliability should not be
severely affected.

\subsubsection{Planetary Nebulae}

Upon visual inspection of the raw catalogs, we find that extended
Planetary Nebulae (PNe) can generate spurious sources in the AllWISE
source catalog that meet our selection criteria. To avoid such
sources, we cross-correlate our catalog with the Strasbourg-ESO
Catalogue of Galactic Planetary Nebulae \citep{acker92}, obtained from
the VizieR Astronomical Server\footnote{\url{vizier.u-strasbg.fr}}. We
conservatively eliminate all sources within twice the radius of a
known PN. If both a radio and an optical diameter are listed for a
given nebula, we assume the larger of the two. The catalog contains
1,142 PNe although many are within the regions close to the GP and GC
removed earlier. This filter removes an additional area of 2 deg$^2$.

\subsubsection{H{\,\sc ii} and Star-Forming Regions}\label{sssec:hii_and_sf_regions}

Similarly to PNe, a large number of sources that meet our selection
criteria are associated with H{\,\sc ii} regions in our Galaxy. Some
of them can be, for example, YSOs and AGB stars, which have similar
colors to AGN in the WISE bands \cite[see,
  e.g.,][]{koenig12,nikutta14}. To avoid such sources we use the
H{\,\sc ii} regions in the \citet{anderson14}\footnote{The catalog
  used was downloaded on 2015 December 28 from
  \url{http://astro.phys.wvu.edu/wise/}} catalog, and again
conservatively eliminate all sources within twice the radius of each
H{\,\sc ii} region in the catalog. The catalog of \citet{anderson14}
contains 8,405 sources and removes an additional 105~deg$^2$ from our
final catalog.

A similar effect is observed near known star-forming regions. To
filter such sources out, we use Lynds' catalogs of Dark and Bright
Nebulae \citep[LDN and LBN, respectively;][]{lynds62,lynds65}. The LDN
catalog only lists the surface area of the nebula, $A_{\rm LDN}$, so
for simplicity we assume a radius $r_{\rm LDN} = \sqrt{A_{\rm
    LDN}/\pi}$. For the LBN catalog we assume a radius equal to half
of the largest diameter measured for the nebula. Instead of using the
conservative approach used before for the PN and H{\,\sc ii} regions,
here we only eliminate sources within the radius of each nebula. This
step eliminates an additional 1,443~deg$^2$ from the final catalog.

\subsubsection{Nearby Galaxies}

Finally, we also consider the possibility of contaminants associated
with well resolved, nearby galaxies. While this is most likely only an
issue for the largest galaxies such as the LMC, SMC and M31, we
conservatively consider all galaxies listed in the Catalog and Atlas
of the Local Volume Galaxies
\citep[LVG;][]{karachentsev13}\footnote{The LVG catalog was retrieved
  on 2015 December 29 from \url{https://www.sao.ru/lv/lvgdb/}}, as
well as all sources in the 2MASS Extended Source Catalog
\citep[XSC;][]{skrutskie06}. As done for several previous stages of
the spatial filtering, we eliminate all sources within twice the
radius of each of the sources in the LVG and 2MASS XSC catalogs. For
the LVG sources we use the Holmberg isophote ($\sim26.5~\rm mag/\rm
arcsec^2$ in $B$-band), while for the 2MASS XSC sources we use the
total radius estimate. There are 1,647,900 sources between the two
catalogs, and this step removes an additional 856~deg$^2$ from the
final AGN catalog.

\subsection{Final Catalog}

The final R90 and C75 AGN catalogs are presented in Tables
\ref{tab:r90_short} and \ref{tab:c75_short} respectively. The R90
catalog contains 4,543,530 sources, while the C75 catalog contains
20,907,127 sources. After applying the spatial filters, the effective
area of the final catalogs is 30,093~deg$^2$. This implies that the
average source density of the R90 catalog is 151 deg$^{-2}$, while the
corresponding value for the C75 catalog is 695 deg$^{-2}$.

\begin{figure}
  \begin{center}
    \plotone{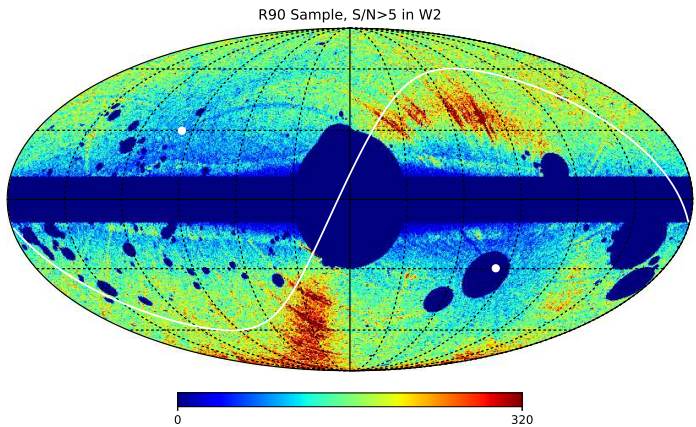}
    \caption{Surface density of sources in the final R90 catalog,
      obtained with HEALPix and displayed using a Mollweide
      projection. The colors display different surface densities in
      units of deg$^{-2}$, as indicated by the color bar at the bottom
      of the Figure. The white solid line shows the plane of the
      Ecliptic, while the solid white dots show the Ecliptic Poles.}
    \label{fg:R90_map}
  \end{center}
\end{figure}

\begin{figure}
  \begin{center}
    \plotone{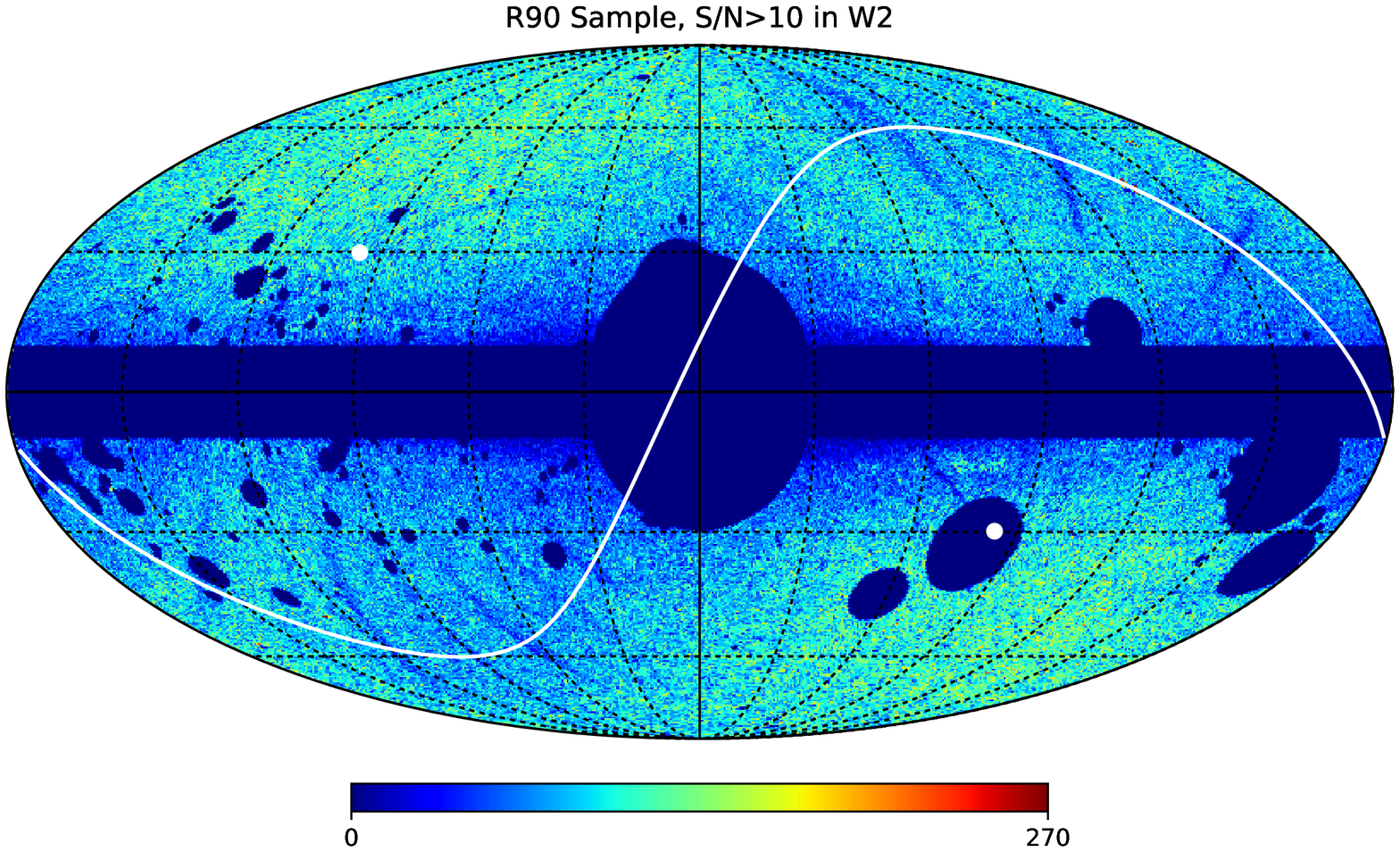}
    \caption{Same as Fig. \ref{fg:R90_map} but limiting the R90
      catalog to only sources detected with $S/N>10$ in W2.}
    \label{fg:R90_SNR10_map}
  \end{center}
\end{figure}

\begin{figure}
  \begin{center}
    \plotone{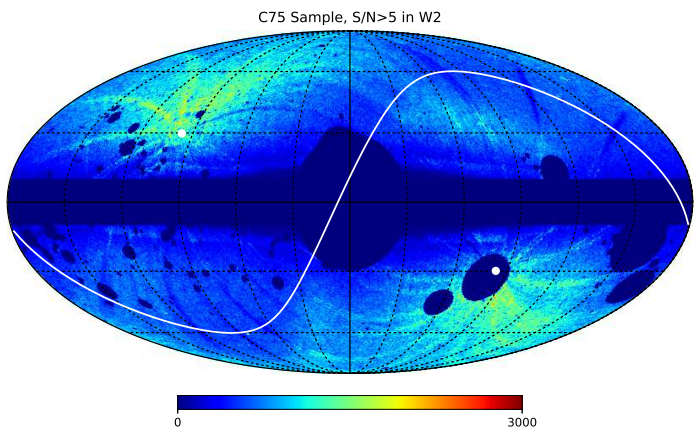}
    \caption{Same as Fig. \ref{fg:R90_map} but for the C75 catalog.}
    \label{fg:C75_map}
  \end{center}
\end{figure}

\begin{figure}
  \begin{center}
    \plotone{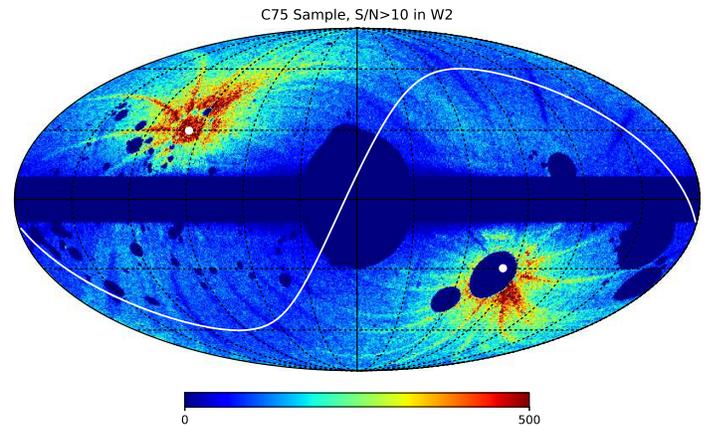}
    \caption{Same as Fig. \ref{fg:C75_map} but limiting the C75
      catalog to only sources detected with $S/N>10$ in W2.}
    \label{fg:C75_SNR10_map}
  \end{center}
\end{figure}

Figures \ref{fg:R90_map}, \ref{fg:R90_SNR10_map}, \ref{fg:C75_map} and
\ref{fg:C75_SNR10_map} show the all-sky source density maps of the R90
and C75 samples, using the Mollweide projection of
HEALPix\footnote{\url{http://healpix.sf.net}. HEALPix functions were
  used through the {\tt{healpy v1.9.1}} package} \citep{gorski05}. The
distribution of sources is not uniform, with both large scale,
smoothly varying structures in the all-sky maps, as well as isolated
high concentration regions.

In the next sections we describe some of these map features and their
origins.

\subsection{Systematic Structures in the All-Sky Map}\label{ssec:systematics}

Most of the systemic features seen in Figures
\ref{fg:R90_map}--\ref{fg:C75_SNR10_map} are related to known
variations in the depth of the WISE survey which result from the WISE
survey
strategy\footnote{\url{http://wise2.ipac.caltech.edu/docs/release/allwise/expsup/sec4\_2.html}}. However,
we also identify additional artifacts introduced by extremely bright
stars, planets and the South Atlantic Anomaly (SAA).

\subsubsection{Smooth Density Gradients Towards the Ecliptic Poles}\label{sssec:smooth_gradient}

The WISE spacecraft is in a polar orbit with a period of 95~min,
taking images every 11~s in the direction perpendicular to the
Earth-Sun line \citep{wright10}. Because the scan lines are along
lines of ecliptic longitude, this survey pattern results in
increasingly denser coverage at higher absolute values of the ecliptic
latitudes, as every scan goes through the ecliptic poles.

The gradients caused by such patterns are most apparent in the all-sky
density maps of the C75 sample (Figs. \ref{fg:C75_map} and
\ref{fg:C75_SNR10_map}) but are much less evident in the R90 maps
(Figs. \ref{fg:C75_map} and \ref{fg:C75_SNR10_map}). The reason for
this difference is that the C75 sample is effectively $S/N$ limited,
implying a source density that increases with survey depth. The R90
selection criteria instead disfavors fainter sources in W2 by
requiring them to be increasingly redder, making it much less
susceptible to differences in survey depth.

Interestingly, however, the highest overdensities are not exactly
coincident with the ecliptic poles (EPs), but are actually located
$\sim 10~\rm deg$ away from the EPs in the direction directly opposite
to the GP. This is most likely due to Galactic dust, which is
increasingly abundant closer to the Galactic Plane \citep[see,
  e.g.,][]{schlegel98}, and could lower the $S/N$ of a given source
either by obscuring its W2 magnitude or by raising the local
background. Hence the location of the highest density regions in the
C75 sample is due to a trade-off between lower dust content and deeper
survey depth.

While these large scale overdensity patterns are mostly dependent on
Ecliptic and Galactic declination, there is also clearly a pattern
that depends on Ecliptic longitude, with features that connect both
Ecliptic Poles. These features are due to the Moon avoidance maneuvers
of the survey strategy which avoids fields highly contaminated by
scattered Moon light. We refer the reader to \citet{wright10} and the
AllWISE Explanatory Supplement for details.

\subsubsection{High Density Regions at the SAA Declinations}\label{sssec:saa}

The SAA is located at intermediate southern Earth latitudes and, as
described by \citet{wright10}, the WISE survey design adopted a
specific approach to deal with the decrease of sensitivity when
nearing this region. The expectation then would be that there should
be no obvious signatures of the SAA in our all-sky source density
maps. This is true for the C75 sample, but is only true for the R90
sample with W2 $S/N>10$. For the R90 sample with W2 $S/N>5$, however,
there are obvious overdensities at such latitudes. This implies that
in these regions there is an excess of red sources near the detection
threshold of the W2 band.

These overdensities are elongated at approximately constant Ecliptic
longitude, suggesting a relation with the survey scanning
pattern. Upon visual inspection of a sample of images in these
regions, we find that they display significant background gradients
due to scattered Moon light. However contamination by Moon-scattered
light is not a unique condition of fields near the SAA, but it is only
the latter that show such a source enhancement.

It is not clear at this point what is the relation between the SAA and
the Moon scattered light that results in an enhancement of red sources
near the detection threshold of the W2 band, and also whether these
sources are real or not, although they are likely related to an excess
of cosmic rays. We hence strongly caution the user when considering
faint sources near the SAA in fields with high Moon background. To aid
in identifying possibly problematic sources, we have added a
{\tt{MOON\_SAA}} flag to the catalog (see Tables \ref{tab:r90_short}
and \ref{tab:c75_short}), which is equal to 1 if the source is at a
declination between -15 and -45 degrees, consistent with the SAA
latitude, has a {\tt{moon\_lev}} flag in W2 equal or greater to 3, and
W2 $S/N\leq 7$.

\subsubsection{Diffraction Spikes}\label{sssec:spikes}

A number of additional overdensities in the R90 and C75 maps can be
associated with spurious sources coincident with diffraction spikes
from bright saturated stars. Diffraction spikes around bright stars
produce a significant number of artifacts, and the AllWISE source
extraction attempts to flag detections that are either contaminated by
or spurious detections of diffraction spikes. The accuracy of the
flagging was limited by the imperfect knowledge of heavily saturated
stars and by changes in the survey sensitivity because of
depth-of-coverage variations around the sky.

Upon visual inspection of these overdense regions, we found that the
algorithm used for the artifact detection sometimes underestimates the
length of the diffraction spikes and hence did not flag a number of
spurious sources. It is, however, only a small fraction of bright
stars for which the length of the diffraction spikes was
underestimated. Inspecting a randomly selected group of the brightest
stars in the WISE catalog, we find that this issue is generally not
observed, implying that the diffraction spike detection algorithm is
generally working properly. It is not clear, however, why the process
would be failing just for a small number of bright stars. It is
possible this issue is due to the inherent difficulty of measuring the
brightness of heavily saturated stars. Additionally, stellar
variability might play a role by effectively varying the length of the
diffraction spikes from image to image. Hence, faint sources near
bright stars should be treated with caution.

\subsubsection{Solar System Planet Residuals}\label{sssec:planets}

Finally, we find a number of spurious sources associated with
residuals left by Solar System planets in the coadded WISE data. While
moving objects are typically suppressed in the coadded images, the
brightest ones, namely Mars, Jupiter and Saturn, can leave residuals
that trigger spurious detections that may persist in the AllWISE
catalog (see the AllWISE Explanatory Supplement for details). Upon
inspection of some of the highest density HEALPix pixels in the
all-sky density maps of the R90 and C75 samples, we find that some
such residuals meet our selection criteria and hence appear in our
final R90 and C75 catalogs. Unlike bright stars, the quick apparent
motion of planets makes it more difficult to deal with in a simple
manner.

Note, however, that spurious sources arising from unflagged artifacts
due to the residuals of Solar System planets and the diffraction
spikes of bright stars result in much higher local surface densities
of AGN. This is also the case for the Moon-contaminated SAA fields, as
well as for fields with PNe, H{\,\sc ii} regions and star-forming
regions outside the areas used to filter the R90 and C75 catalogs in
\S\ref{ssec:spatial_filters}. Considering this, we include for each
source in the final R90 and C75 catalogs the surface density of the
HEALPix pixel that contains it ($\Sigma_{\rm pix}$) so the user can
decide how to best deal with the described artifacts.

\begin{figure}
  \begin{center}
    \plotone{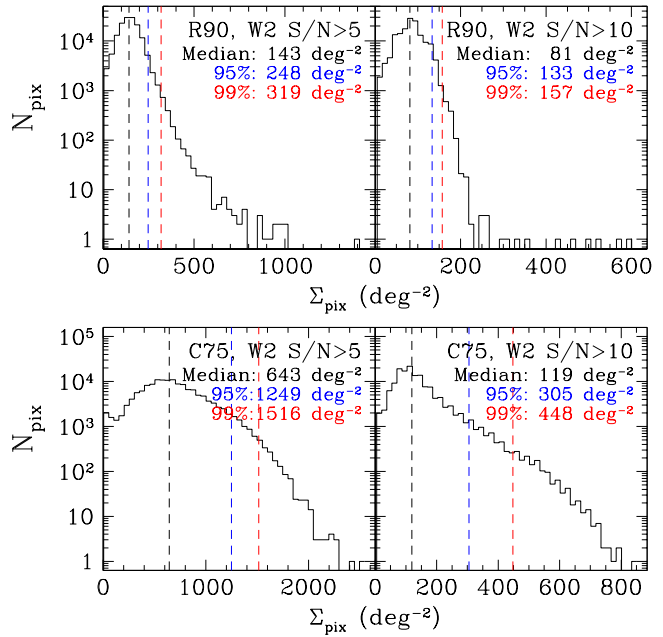}
    \caption{Distribution of $\Sigma_{\rm pix}$, the AGN candidate
      surface density in HEALPix pixels ({\tt{NSIDE}}=2$^7$)
      containing at least one object. The area of each pixel is
      0.21~deg$^{2}$. The dashed black line shows the median of the
      distribution, while the blue and red lines show the 95$^{\rm
        th}$ and 99$^{\rm th}$ percentiles. The highest pixel
      densities might be indicative of the contaminants discussed in
      \S\ref{ssec:systematics}.}
    \label{fg:hist_pixels}
  \end{center}
\end{figure}

Figure \ref{fg:hist_pixels} shows the distribution of $\Sigma_{\rm
  pix}$ for the R90 and C75 catalogs, including their restricted
versions requiring $S/N>10$ in W2. Each Figure also shows the median,
95${\rm th}$ and 99${\rm th}$ percentile of the distributions, which
are listed in Table \ref{tab:healpix_dens}.

\section{Highly Variable Objects}\label{sec:agn_var}

The AllWISE catalog classifies sources according to their probability
of variability, which is determined from forced photometry at the
individual frame level (see the AllWISE Data Release Explanatory
Supplement for details). The classification is done per band,
assigning a number ranging from 0 through 9 going from least to most
probable for variability.

Here, we focus on the subset of sources from the R90 catalog that are
most likely to be real variables. Specifically, we select all objects
that have a variability flag of 9 in both the W1 and W2 bands. Only
687 such sources, corresponding to 0.015\% of the R90 catalog, match
this criterion. Given the survey design, there are two natural
cadences for the WISE data: the cadence of $\sim 3~\rm hrs$ that
corresponds to twice the orbital period of the satellite, and the
cadence of about 6 months set by half of the orbital period of the
Earth around the Sun. For every region in the sky, WISE obtained at
least 8 images (with coverage increasing with distance to the
Ecliptic) separated in time by the shorter cadence, and then returned
to the same region at least once more with a time separation of 6
months.

The sensitivity to short variability timescales means that a
significant number of these 687 sources are likely to be blazars. To
assess the fraction of these sources that are blazars we use the FIRST
survey \citep{becker95}. While in general cross-matching radio surveys
with surveys at other wavelengths can be quite challenging due to
highly extended radio structures that necessitate sophisticated
approaches \citep{devries06}, blazars avoid this issue since they
contain compact, beamed radio cores. Therefore, we simply use a
5\arcsec\ matching radius to find counterparts between our highly
variable R90 catalog sources and FIRST sources through the official
FIRST Catalog Search
tool\footnote{\url{http://sundog.stsci.edu/cgi-bin/searchfirst}}. Of
the 687 mid-IR variable sources, 207 are within the FIRST footprint,
and 162 (78\%) are detected by FIRST. The remaining 45 objects (22\%)
are not detected by FIRST and are therefore unlikely to be blazars. In
the next section we discuss the spectra of some of those highly
variable R90 AGN candidates, while in \S\ref{ssec:w1428} we focus on
one of these objects for which new spectroscopy reveals that the
source is a changing look quasar, transitioning from a type 1 to a
type 2 AGN. The lightcurves of the variable sample will be discussed
in detail in \citet{assef16}.

For completeness, we also cross-match our highly variable R90 sources
with the source catalog of the NVSS survey \citep{condon98}. We
obtained the NVSS source catalog through the VizieR Astronomical
Server. We find that 411 of the highly variable R90 sources are within
22.5\arcsec\ (HWHM of the NVSS beam) of an NVSS source, 251 of which
are outside of the FIRST footprint. There are a total of 150 highly
variable AGN within the NVSS footprint (i.e., with declination
$>$--40~deg) but without an NVSS source within 22.5\arcsec. Of these,
103 are outside the FIRST footprint. Given the somewhat shallower
depth of the NVSS survey as well as the very large beam size, these
results are somewhat harder to interpret, and hence we focus the
discussion of the following sections only on those objects within the
FIRST survey footprint.

\subsection{Optical Spectroscopy}\label{ssec:var_specs}

Of the 687 highly variable sources, 136 have optical spectra in the
Sloan Digital Sky Survey Data Release 12 \citep{alam15}. Of these, 132
are within the FIRST survey footprint, and 103 have measured fluxes at
1.4~GHz. This implies that 29 out of the 45 non-radio, highly variable
AGN have optical spectra from SDSS. Their spectroscopic redshifts and
classifications are listed in Table \ref{tab:var_rq_specs}. We also
add spectroscopic redshifts and classifications for four more objects
from SIMBAD. For the SDSS objects classified as stars as well as those
with significant warnings from the SDSS pipeline we show the SIMBAD
classification instead. Finally, we also add a photometric redshift
and classification from SIMBAD for WISEA J150954.94+203619.6. Of the
33 non-radio, highly variable WISE AGN candidates we find that 19 are
classified as type 1 AGN (either QSO or Seyfert 1), six are classified
as ``Galaxy AGN'' (meaning they have narrow-emission lines
characteristic of type 2 AGN), one is classified as a possible AGN,
four are classified as galaxies, and three are classified as
stars. Upon inspection of the spectra of the four objects classified
as galaxies, we find that their H$\alpha$ emission lines have
significantly broadened bases, suggesting an important AGN
contribution. Of the three sources classified as stars, two are
classified as carbon stars. These cool giant stars can produce
significant amounts of dust. For the remaining object classified as a
star, WISEA J163518.38+580854.6, no further information on its nature
is provided by SIMBAD. However, this object is likely associated to
the {\it{ROSAT}} X-ray source 1RXS J163518.7+580856 located only
2.78\arcsec\ away, implying it may be an unrecognized quasar. Note
that if the 90\% reliability of the R90 sample were to hold for this
subgroup of highly variable mid-IR AGN candidates, we would have
expected about three of the 33 sources to be contaminants, consistent
with the number of Galactic sources found if all four targets
classified as galaxies host AGN activity.

As the SDSS targeting criteria is biased towards unobscured AGN, we
complement this sample with long-slit optical spectroscopic
observations obtained for five additional highly variable AGN
candidates within the SDSS survey footprint but without SDSS
spectra. The observations were carried out on the night of UT 2016
February 6 using the DBSP optical spectrograph at the Palomar
Observatory 200-inch telescope. We used the D55 dichroic with the 600
lines/mm grating (4000\AA\ blaze) on the blue arm and the 316 lines/mm
grating (7500\AA\ blaze) on the red arm. The slit used had a width of
1.5\arcsec. Due to scheduling constraints, most of the targets
selected were in regions close to the Galactic Plane. Reductions were
carried out in a standard manner using
IRAF\footnote{\url{http://iraf.noao.edu}}.

Table \ref{tab:var_rq_specs_palomar} shows the results of these
observations. We first observed two sources that were not detected in
the FIRST survey, as per the sources listed in Table
\ref{tab:var_rq_specs}. We find that both are AGN. WISEA
J015858.48+011507.6 has a spectrum consistent with a type 2 AGN at
$z=0.184$, with high [O\,{\sc iii}]/H$\beta$ and [N\,{\sc
    ii}]/H$\alpha$ ratios and clear detection of the high excitation
[Ne\,{\sc v}] line. WISEA J101536.17+221048.9 shows broad emission
lines and a continuum consistent with a reddened type 1 AGN at
$z=0.555$.

Additionally, we observed two sources that are well detected in the
FIRST survey but are within the 25\% faintest radio fluxes. WISEA
J090931.55-011233.3 ($F_{1.4~\rm GHz} = 15.66\pm 0.15~\rm mJy$) has a
spectrum consistent with a type 2 AGN at $z=0.201$, although with a
red continuum and unusually low equivalent width emission lines. WISEA
J095528.76+572837.2 ($F_{1.4~\rm GHz} = 21.66\pm 0.15~\rm mJy$) shows
a featureless continuum consistent with a blazar.

Finally, we observed one source, WISEA J051939.78+160044.0, outside of
the FIRST radio survey areal coverage. This source is a Galactic
cataclismic variable (CV), likely associated with the {\it{ROSAT}}
source 1RXS J051939.7+160042 which is offset by only
2\arcsec\ according to
SIMBAD\footnote{\url{http://simbad.u-strasbg.fr/simbad/}}. This source
is in the vicinity of the Orion Nebula and only 12 deg away from the
GP, so its Galactic nature is reasonably expected.

\subsection{WISEA J142846.71+172353.1: A Changing Look Quasar}\label{ssec:w1428}

\begin{figure}
  \begin{center}
    \plotone{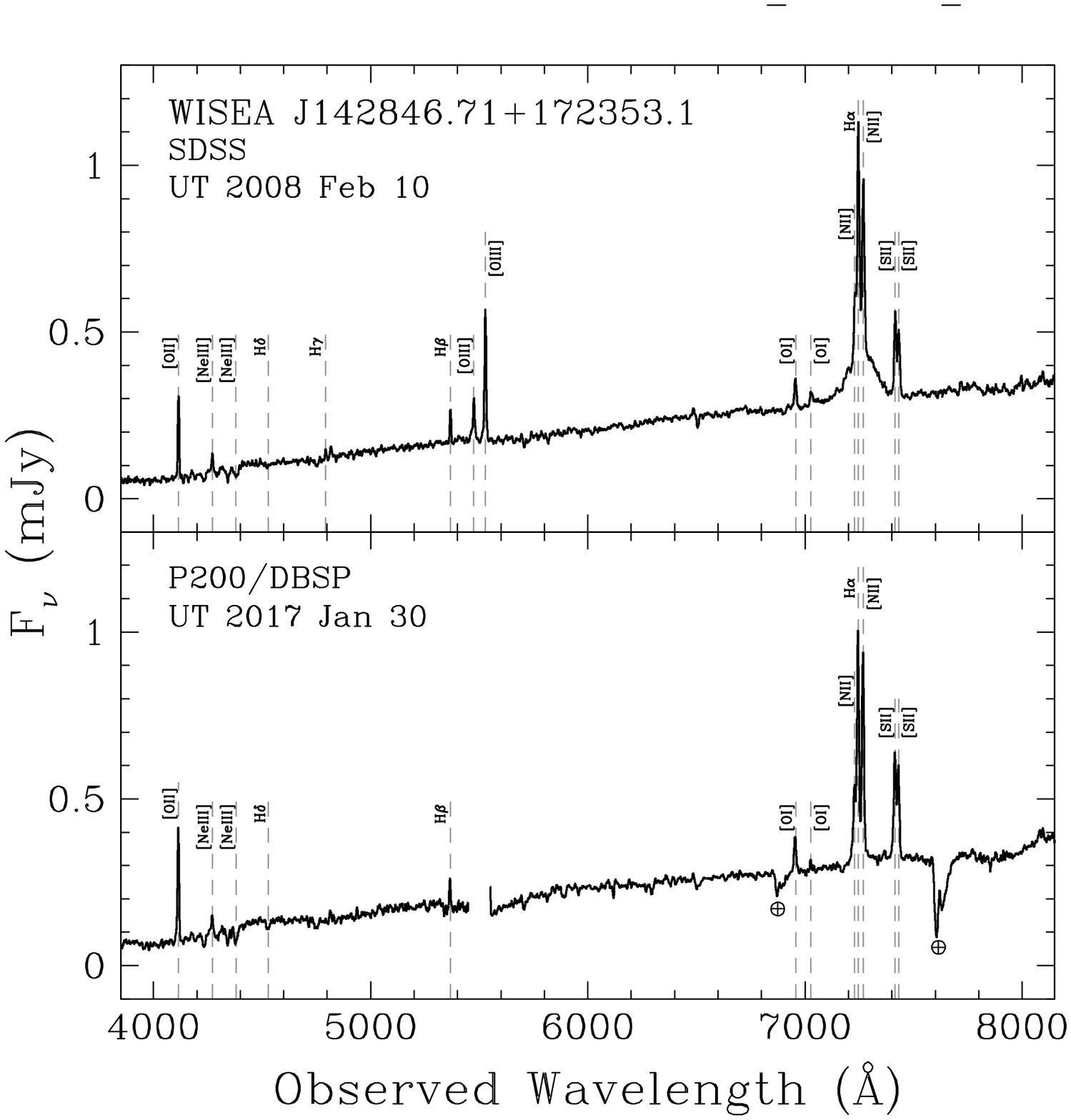}
    \caption{Optical spectroscopic observations of WISEA
      J142846.71+172353.1. The top panel shows the earlier spectrum
      obtained by SDSS while the bottom panel shows a recently
      obtained spectrum with the DBSP instrument at the Palomar
      Observatory 200-inch telescope. Note that for the latter we have
      not corrected for telluric absorption (i.e., A-band at about
      7600--7630\AA\ and B-band at about 6860--6890\AA). The broad
      component of H$\alpha$ observed in the SDSS spectrum is missing
      in the recent Palomar observations.}
    \label{fg:w1428_specs}
  \end{center}
\end{figure}

We obtained new spectroscopic observations for WISEA
J142846.71+172353.1, classified as a broadline QSO by SDSS (see Table
\ref{tab:var_rq_specs}), to study its spectral evolution in the face
of the strong WISE variability. Observations were obtained on the
night of UT 2017 January 30 (MJD 57783) with the DBSP optical
spectrograph at the Palomar Observatory 200-inch telescope using the
same setup described in the previous section. Figure
\ref{fg:w1428_specs} shows the resulting spectrum, as well as the
earlier SDSS spectrum obtained on the night of UT 2008 February 10
(MJD 54506).

\begin{figure}
  \begin{center}
    \plotone{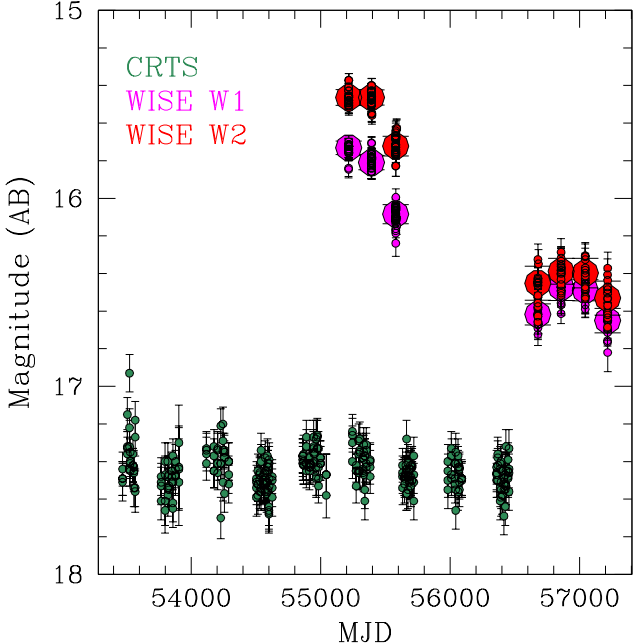}
    \caption{Light curve of WISEA J142846.71+172353.1 in the optical
      from CRTS (green points), and in the WISE bands W1 (purple
      points) and W2 (red points). All magnitudes are shown in the AB
      system to limit the dynamic range of the vertical axis. For W1
      and W2, the small circles show the individual frame photometry
      in the AllWISE and NEOWISE-R surveys. The large circles show the
      median of each epoch.}
    \label{fg:w1428_lc}
  \end{center}
\end{figure}

A decade ago, the source exhibited a clear broad component to the
H$\alpha$ emission line that is not present in 2017. However, neither
spectrum shows a broad H$\beta$ component, indicating that the source
transitioned from an intermediate-type AGN at the time of the SDSS
observations to a type 2 AGN at the time of our Palomar
observations. Figure \ref{fg:w1428_lc} shows the W1 and W2 light curve
of this source. We include the latest publicly available W1 and W2
data from NEOWISE-R \citep{mainzer14}. Additionally, we include its
optical light curve from the Catalina Real-Time Transient
Survey\footnote{\url{http://nesssi.cacr.caltech.edu/DataRelease/}}
\citep[CRTS;][]{drake09}, retrieved from the Catalina Surveys Data
Release 2. Between the WISE and NEOWISE-R epochs the source dimmed by
approximately 1 mag in both W1 and W2, with further, lower amplitude
variability observed between the epochs of each survey
independently. In contrast, there is no strong optical variability
observed by CRTS. This is consistent with both spectroscopic
classifications, as in both an intermediate-type and a type 2 AGN the
optical emission is dominated by the host galaxy. The large drop in
mid-IR fluxes suggests that the change in spectroscopic classification
is most likely due to a decrease in accretion rate rather than a
change in obscuration.

\section{Comparisons with Other Surveys}\label{sec:other_cats}

As mentioned earlier, a number of AGN catalogs already exist in the
literature that are similarly large to ours. In this section we
compare with a number of them in order to characterize how they differ
from our R90 and C75 catalogs, as well as assess what the relative
completeness and biases are. We consider both mid-IR selected catalogs
as well as catalogs selected in different wavelength ranges and,
hence, affected by very different systematics.

\subsection{\citet{secrest15} WISE AGN Catalog}

Recently, \citet{secrest15} presented an all-sky AGN catalog selected
purely on their WISE colors, based on the selection criteria of
\citet{mateos12}. These criteria require a detection in the W3 band.
A13 showed these criteria to be less reliable than the R90 criterion
at faint W2 magnitudes, though they are equally reliable at brighter
magnitudes. The trade off is that the \citet{mateos12} selection
criteria are more complete at faint W2 magnitudes than the R90
selection. As noted by \citet{jarrett11}, requiring a detection in W3
can be particularly useful near the high survey coverage areas at the
ecliptic poles, where the W1 and W2 are confusion limited, and the
lower source density W3 band provides robust photometry. However, the
lower sensitivity of the W3 band restricts the AGN sample size created
using the selection criteria of \citet{mateos12}. Additionally,
\citet{secrest15} required a 5$\sigma$ detection in all three WISE
bands, creating a robust catalog at the cost of decreasing the number
of sources selected. Indeed, the full catalog presented by
\citet{secrest15} consists of 1,354,775 sources, and once we apply the
same spatial filters as described in \S\ref{ssec:spatial_filters}, the
catalog is reduced by 15\% to 1,140,022 sources. This is roughly 25\%
of the sources in our R90 catalog.

Comparing the catalogs we find that only 50,877 (4.5\%) of the
\citet{secrest15} AGN candidates after applying the spatial filtering
are not contained in our R90 sample. This number is further reduced to
42,565 (3.7\%) when we eliminate objects that do not meet the
additional requirements we imposed in \S\ref{sec:agn_cats}. These are
likely real AGN that fall outside the R90 selection criterion. Our R90
sample hence recovers the great majority of the objects selected as
AGN by \citet{secrest15} but contains approximately four times more
sources, making it a much more complete AGN sample with comparable
reliability. Comparing to our C75 catalog instead, we find that only
17,284 (1.5\%) of the \citet{secrest15} AGN candidates, after applying
the spatial filtering and additional requirements, are not contained
in it.

\subsection{Match to the updated XDQSOz catalog}

As another comparison of this new WISE AGN catalog to other large,
multi-wavelength quasar catalogs, we matched the R90 sample to the
updated extreme deconvolution quasar catalog presented by
\citet{dipompeo15}.  \cite{bovy11} originally developed and applied an
extreme deconvolution technique to build a quasar catalog (XDQSO) from
all point sources in Data Release 8 of the Sloan Digital Sky Survey
\citep{aihara11}.  \citet{bovy12} added photometric redshift
information, UV photometry, and near-IR photometry to produce the
XDQSOz.  That catalog was further updated by \citet{dipompeo15} to
incorporate public all-sky WISE photometric data, which improves both
quasar likelihood assessments and photometric redshifts; we refer to
the updated catalog as the uXDQSOz.  The uXDQSOz identifies 5,537,436
potential quasars with probability $P_{\rm QSO} \geq 0.2$, or
3,874,639 quasars weighted by probability.

As before, we apply our spatial filtering procedure to the uXDQSOz,
which reduces the number of uXDQSOz sources within our R90 footprint
to 4,105,027.  We then match the two catalogs with
TOPCAT\footnote{Available at
  \url{http://www.star.bristol.ac.uk/~mbt/topcat/}.} \citep{taylor05}
using a matching radius of $4.5\arcsec$ determined using single
matches only (i.e., closest pairs); above this threshold, chance
coincidences start to become significant.  We obtain 631,662 matches,
of which $>99\%$ are single matches.  This represents just 15.4\%\ of
uXDQSOz sources within the area under consideration.  However,
83.6\%\ of the matched R90 sources have $P_{\rm QSO} \geq 0.9$, as
compared to just 42.1\% of the uXDQSOz within this area. These
percentages become even more extreme if we consider that 80.7\%\ of
the matched sources have $P_{\rm QSO} \geq 0.95$ while only 37.3\%\ of
the entire uXDQSOz catalog within the R90 footprint has this very high
likelihood of being quasars. Many of the uXDQSOz high-likelihood
quasars not identified by the WISE color selection are at higher
redshifts ($z \gtrsim 3$), where the observed mid-IR colors become
bluer (e.g., Fig. 1 of A13). Conversely, WISE identifies robust quasar
candidates across most of the sky, whereas the uXDQSOz is restricted
to the SDSS footprint.  Furthermore, WISE identifies obscured quasars,
most of which would be lost by the initial requirement of the extreme
deconvolution quasar samples that the target be unresolved in SDSS
optical imaging.

\subsection{The SDSS Quasar Catalog}

We compare our WISE-selected AGN catalogs with the latest edition of
the SDSS quasar catalog, based on the 12$^{\rm th}$ data release of
the survey \citep{paris17}. Because SDSS is an optical survey, the
DR12 quasar catalog preferentially contains unobscured AGN. We refer
the reader to \citet{ross12} and \cite{paris17} for the exact details
of sample selection.

The SDSS DR12 quasar catalog contains 297,301 spectroscopically
confirmed AGN. We find that 209,758 (70\%) of these sources have a
counterpart in the AllWISE catalog within a matching radius of
2\arcsec. This number reflects the fact that SDSS targets significant
numbers of quasars that are fainter than the WISE detection
limits. However, this fraction is higher than the 64\% AllWISE matches
reported by \citet{paris17} for the same matching radius (190,408
sources within the entire catalog). The source of the discrepancy is
currently unknown but likely relates to additional quality flags
applied by \citet{paris17} on the WISE photometry. Of these 209,758
WISE matches, 158,356 (75\%) meet the data quality requirements used
to build the main sample from which the R90 and C75 samples were
generated in \S\ref{sec:agn_cats} (i.e., WISE point sources, not
flagged as either artifacts or affected by artifacts, fainter than the
saturation limits in W1 and W2, and with W2 $S/N>5$) and are within
the area allowed by the spatial filters applied in
\S\ref{ssec:spatial_filters}.

\begin{figure}
  \begin{center}
    \plotone{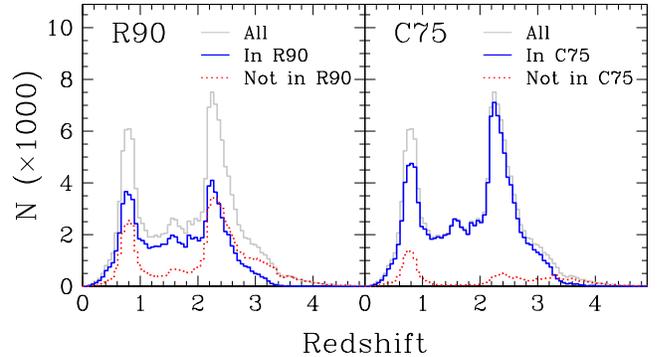}
    \caption{{\it{(Left)}} Redshift distribution of SDSS DR12 quasars
      with matches in the AllWISE catalog that pass the requirements
      outlined in \S\ref{sec:agn_cats}. The gray histograms show the
      distribution of quasars, while the blue (red) lines shows those
      found (not found) within the R90 catalog. {\it{(Right)}} Same as
      the left panel but comparing to the C75 catalog.}
    \label{fg:sdss_redshift_comparison}
  \end{center}
\end{figure}

\begin{figure}
  \begin{center}
    \plotone{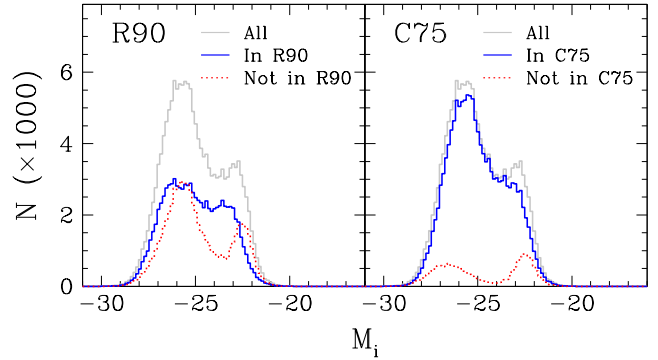}
    \caption{Absolute i--band magnitude distribution of SDSS DR12
      quasars with matches in the AllWISE catalog that pass the
      requirements outlined in \S\ref{sec:agn_cats}. Lines and panels
      have the same definition as in
      Fig. \ref{fg:sdss_redshift_comparison}.}
    \label{fg:sdss_Mi_comparison}
  \end{center}
\end{figure}

Cross-matching with our WISE-selected AGN catalogs, we find that
90,326 (30\%) of the SDSS AGN are in the R90 sample, and 138,410
(47\%) are in the C75 sample. This means that 57\% and 87\% of the
objects in the SDSS DR12 quasar catalog with WISE matches that pass
the data quality and spatial filter requirements of our main sample
are found in the R90 and C75 samples respectively. Note that the
completeness is higher than expected for the C75 sample, implying that
SDSS misses a fraction of the WISE-detected AGN used to calibrate the
selection in \S\ref{sec:agn_selection}. The fraction of SDSS AGN
missed by the R90 and C75 catalogs is not random though, but rather
depends significantly on other parameters. Figure
\ref{fg:sdss_redshift_comparison} shows the redshift distribution of
SDSS quasars recovered by the R90 (left panel) and C75 (right panel)
criteria, as well as of those with matches in the AllWISE catalogs
that meet all the requirements of \S\ref{sec:agn_cats} but were not
recovered by the respective selection criteria. The redshift
distribution of the SDSS quasar catalog is triple peaked. The peaks at
$z\sim 0.8$ and $z\sim 1.6$ are due to degeneracies in the SDSS
color-redshift space \citep{ross12,paris17}, while the peak at $z\sim
2.3$ is mostly related to the Baryon Oscillation Spectroscopic Survey
experiment \citep[BOSS;][]{dawson13} which primarily targeted $2.15\le
z \le 3.5$ quasars. The R90 criterion recovers SDSS AGN with a higher
efficiency in the $1\lesssim z \lesssim 2$ range. At $z\gtrsim 2$, the
W1--W2 color of unobscured AGN starts becoming progressively bluer
(see, e.g., Fig. 1 of A13), while the R90 color cut becomes
progressively redder for fainter W2 magnitudes (see
Fig. \ref{fg:agn_sel} and eqn. [\ref{eq:rel_sel}]). At $z\lesssim 1$
the recovered fraction is somewhat lower most likely due to missing
the less luminous AGN that will have a higher host-galaxy contribution
to the total luminosity. The stellar emission of those objects will
make them have somewhat bluer W1--W2 colors that are missed at faint
W2 fluxes by the R90 selection criterion. The C75 criterion has a much
higher recovery rate at all redshifts, containing a very large
fraction of all the SDSS quasars in the $1\lesssim z \lesssim 3$
redshift range. At $z>3$ the W1--W2 colors of unobscured AGN become
too blue to be selected by the C75 criterion, in part due to the
contribution of the broad H$\alpha$ emission line to the W1 band
\citep[][A13]{assef10}. While the $z\lesssim 1$ the recovery rate is
also much higher than for the R90 criterion, the lower efficiency
compared to higher redshifts is most likely also due to the
host-galaxy contamination discussed above for the R90
selection. Figure \ref{fg:sdss_Mi_comparison} shows a similar
comparison but for the absolute $i$-band magnitudes instead of
redshift. As for the redshift distribution, the recovery efficiency of
the R90 criterion is highest for intermediate luminosities, while
being lower at the bright and faint ends, which primarily correspond
to the highest and lowest ends of the redshift
distribution. Similarly, the C75 criterion has a higher recovery for
$M_i\lesssim -24$ which corresponds to the highest redshift ranges.

\subsection{Second {\it{ROSAT}} All-Sky Survey}

The {\it ROSAT} X-ray satellite scanned the entire sky between June
1990 and August 1991 in the $0.1 - 2.4\, {\rm keV}$ energy band,
making it the most sensitive all-sky high-energy survey to date and
the best suited for comparing X-ray and WISE all-sky mid-IR AGN
selection.  For these scanning mode observations, the {\it ROSAT} beam
has a FWHM of $\sim 30\arcsec$.  The second release of these
observations, presented as the Second {\it ROSAT} All-Sky Survey
\citep[2RXS;][]{boller16}, includes 135,118 X-ray sources down to a
likelihood threshold of 6.5 (i.e., {\sc EXI\_ML $\geq 6.5$}), where
the catalog is expected to contain about 30\% spurious detections.
Adopting a more conservative likelihood threshold of {\sc EXI\_ML
  $\geq 9$}, the catalog contains 74,453 sources with an expected 5\%
spurious fraction.  The flux limit of 2RXS corresponds to $\sim
10^{-13}\, {\rm erg}\, {\rm cm}^{-2}\, {\rm s}^{-1}$.

\begin{figure}
  \begin{center}
    \plotone{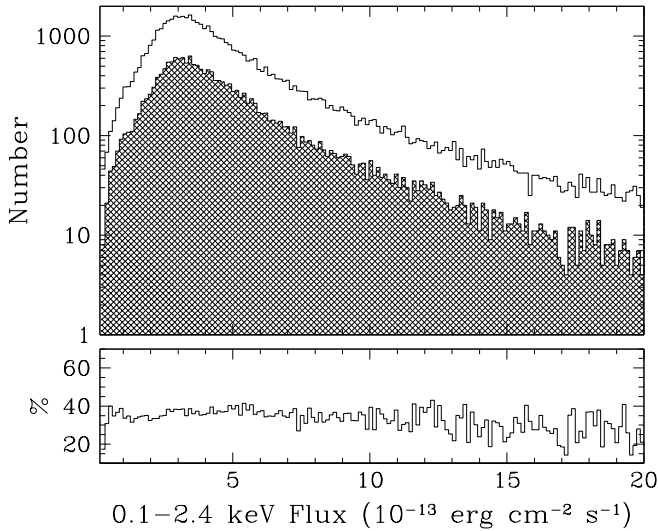}
    \caption{Open histogram in top panel shows the distribution of
      0.1-2.4~keV X-ray fluxes for the 2RXS, considering the 51,973
      sources with {\sc EXI\_ML $\geq 9$} ($\sim 5\%$ spurious
      fraction) and subject to the spatial filtering discussed in
      \S\ref{ssec:spatial_filters} to avoid the Galactic plane and
      other areas subject to elevated levels of false positive
      mid-IR-selected AGN candidates.  Fluxes have been calculated by
      multiplying the count rate by $1.08 \times 10^{-11}\, {\rm
        erg}\, {\rm cm}^{-2}$, which assumes an X-ray power-law model
      \citep[c.f.,][]{boller16}.  The filled histogram shows the X-ray
      flux distribution of this subset of 2RXS sources with
      mid-IR-selected AGN candidate counterparts.  The bottom panel
      shows the matched fraction as a function of X-ray flux.  The
      fraction is relatively constant at a value slightly below 40\%,
      except at the highest fluxes where the fraction dips.}
    \label{fg:rosat_frac}
    \end{center}
\end{figure}

To match the R90 quasar catalog to 2RXS, we begin by applying the
spatial filtering procedure described in \S\ref{ssec:spatial_filters},
which reduces the number of 2RXS sources under consideration to 51,973
for the more conservative likelihood threshold.  Using TOPCAT, we
match the filtered 2RXS catalog to the R90 catalog using a matching
radius of $36\arcsec$ and allowing for multiple matches.  This radius
was determined using single matches only (i.e., closest pairs); above
this threshold, chance coincidences start to dominate.
\cite{boller16} use a similar value ($40\arcsec$) when matching 2RXS
to the Tycho-2 catalog (\citealp{hog98}).  We obtain 18,241 matches,
corresponding to 35.1\%\ of the X-ray sources, but only 0.4\% of the
R90 sources.  Figure \ref{fg:rosat_frac} shows the distribution of
source X-ray fluxes for the spatially-filtered conservative likelihood
threshold, as well as the distribution and fraction matched to WISE
AGN candidates.  We see that the bulk of the 2RXS sources have fluxes
of a ${\rm few} \times 10^{-13}\, {\rm erg}\, {\rm cm}^{-2}\, {\rm
  s}^{-1}$ and that the fraction with WISE AGN candidate counterparts
varies only slightly with flux, dropping at the highest fluxes.  The
64.9\%\ of 2RXS sources not associated with R90 sources likely
represent a combination of spurious X-ray sources, Galactic X-ray
sources, and galaxy clusters.  For example, considering more than 2000
high Galactic latitude ($\left| b \right| > 30^\circ$) {\it ROSAT}
sources with X-ray fluxes $\gtrsim 10\times$ the detection threshold
from the {\it ROSAT} Bright Survey, \citet{schwope00} show that
approximately half the X-ray sources are Galactic, with the remaining
split approximately 2:1 between X-ray AGN and galaxy clusters (i.e.,
only 32.3\%\ of bright {\it ROSAT} sources are AGN).  Assuming no
dramatic changes as one considers sources closer to the {\it ROSAT}
detection threshold, the results here suggest that the vast majority
of {\it ROSAT} AGN are identified by the R90 selection
criterion.

The intersection of X-ray-selected and mid-IR-selected AGN has been
addressed multiple times previously (i.e., \citealp{gorjian08,
  hickox09, wilkes09, eckart10, donley12, stern12, mendez13}).  Mid-IR
AGN selection requires the AGN component to dominate over the host
galaxy SED in the observed mid-IR bands, restricting such selection to
more luminous AGN (in the quasar regime), albeit with enhanced
sensitivity to even heavily obscured AGN compared to optical quasar
selection.  X-ray selection has much less emission to contend with
from stellar-related processes, making X-ray selection sensitive to
lower luminosity AGN, reaching into the Seyfert regime.  However, most
sensitive wide-field X-ray surveys to date are in the lower energy, or
soft X-ray regime ($< 10$~keV), making them susceptible to absorption
and thus less comprehensive for obscured AGN selection.  This is
particularly true for {\it ROSAT}, with its high-energy cut-off at
2.4~keV.  Illustrating the luminosity dependence, \citet{eckart10}
compares X-ray and mid-IR selection of several hundred AGN and AGN
candidates using data from six relatively deep fields observed by {\it
  Chandra} and {\it Spitzer}.  While $> 80\%$ of X-ray AGN with $L_X >
10^{44}\, {\rm erg}\, {\rm s}^{-1}$ are selected using the
\citet{stern05} {\it Spitzer} mid-IR AGN selection criteria, this
fraction drops monotonically with X-ray luminosity, such that only
36\%\ of sources with $L_X < 10^{43}\, {\rm erg}\, {\rm s}^{-1}$ are
selected by the {\it Spitzer} mid-IR AGN selection criteria.
Therefore, the {\it ROSAT}-detected AGN not selected by the R90
criterion are expected to primarily be lower luminosity AGN, while the
WISE-selected AGN not detected by {\it ROSAT} are likely to be
luminous quasars below the {\it ROSAT} detection threshold including
obscured quasars.

We note that recently, \citet{salvato17} has presented a catalog of
AllWISE counterparts to the 2RXS catalog sources. Instead of simply
relying on positional proximity as done above, \citet{salvato17} uses
a Bayesian matching algorithm that considers the astrometric
information of the sources, as well as a prior on the color and
magnitude of the AllWISE sources determined empirically from the
cross-match between the AllWISE catalog and the 3XMM-DR5 catalog of
X-ray sources \citep{rosen16}, which is considerably deeper than the
2RXS catalog. \citet{salvato17} finds at least one AllWISE counterpart
to 48,416 2RXS sources that pass the spatial filters described in
\S\ref{ssec:spatial_filters} and that have {\sc EXI\_ML $\geq
  9$}. Because of the nature of their approach, the catalog does not
differentiate between AGN and non-AGN sources. We find that of the
best-matched AllWISE source to those 48,416 X-ray sources \citep[i.e.,
  those with {\tt{match\_flag=1}}; see][for details]{salvato17},
19,109 (39.5\%) are in the R90 catalog. Of the 29,307 sources not in
the R90 catalog, we find that only 15,777 meet the additional
requirements we imposed in \S\ref{sec:agn_cats}, and are likely a
combination of Galactic sources, galaxy clusters and low luminosity
AGN as discussed earlier, as well as some chance
alignments. Specifically, if we compare for these 15,777 sources their
{\tt{p\_any}} values, defined by \citet{salvato17} as the probability
that any of their AllWISE associations to a 2RXS sources is the
correct one, we find that 40\% have {\tt{p\_any}} above 0.8 and 32\%
have {\tt{p\_any}} below 0.2. Instead for the 19,109 sources that are
in the R90 catalog we find that 80\% have {\tt{p\_any}} above 0.8 and
only 2\% have {\tt{p\_any}} below 0.2, suggesting a significantly
lower fraction of chance alignments.

\subsubsection{Quasar Triplets with ROSAT Counterparts}

Of the 18,241 matches between the R90 and filtered 2RXS catalogs
discussed above, 17,217 (94.4\%) are single matches.  The remainder
are multiple matches, where two or more R90 AGN candidates are within
36\arcsec\ of a 2RXS X-ray source. Multiple quasar systems are
extremely rare, with only a few confirmed cases reported in the
literature \citep{djorgovski07,farina13,hennawi15}. We consider this
sample in greater detail next, as it has the potential to identify
galaxy clusters based on an overdensity of AGN. In particular, since
the AGN are more common in distant galaxy clusters, with a rate that
vastly outpaces their field evolution \citep[e.g.,][]{galametz12,
  martini13}, this could be a promising method to identify distant ($z
> 1$) X-ray emitting galaxy clusters, and illustrates just one of the
multitude of new studies enabled by this WISE AGN catalog.

Most of the multiple matches correspond to two R90 sources matched to
a single 2RXS source, but there are 33 cases of a single X-ray source
having three R90 sources within the matching radius, one case of a
single X-ray source matching with four R90 sources
(2RXS~J094004.6+122047), and one extreme case of a single X-ray source
having eight R90 sources within $36\arcsec$ (2RXS~J002057.4$-$194632).
The most extreme overdensities prove to be spurious, where the octet
is associated with an excess of sources in a field affected by both
the SAA and the Moon (c.f., see \S\ref{sssec:saa}), and the quartet is
associated with diffraction spikes from the well-studied, IR-bright
carbon star IRC+10216 (also known as CW~Leonis), which is the
brightest 5$\mu$m source in the sky outside the Solar System (see
\S\ref{sssec:spikes}).  Considering the 33 triplets, several also seem
to be affected by elevated noise associated with SAA and lunar
passages.  \S\ref{sssec:saa} shows that the SAA and lunar
contamination is significantly less problematic if one only considers
W2 sources detected at $\geq 10\sigma$.  One draconian, but effective,
method to avoid contamination is therefore to require bright mid-IR
AGN candidates.  Adopting a photometric limit of $W2 \leq 15.05$,
roughly corresponding to the $10\sigma$ threshold at the shallowest
regions of the WISE survey \citep{stern12}, reduces the sample of
triplets to the four 2RXS sources listed in Table
\ref{tab:rosat_triplets}.

\begin{figure*}
  \begin{center}
    \includegraphics[width=0.7\textwidth]{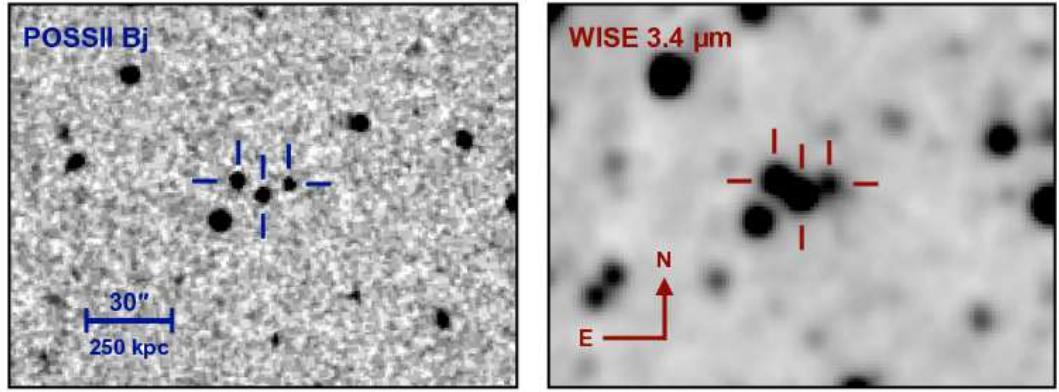}
    \caption{Images of 2RXS~J150158.6+691029 at optical wavelengths
      ($Bj$ band, from POSS~II; left) and mid-IR wavelengths (WISE W1
      band; right).  Images are $\sim 3\arcmin\, \times 2.5\arcmin$,
      with North up and East to the left.  Highlighted are the three
      WISE-selected AGN candidates within the 36\arcsec\ of the {\it
        ROSAT} source.}
    \label{fg:triplet_image}
  \end{center}
\end{figure*}

\begin{figure}
  \begin{center}
    \plotone{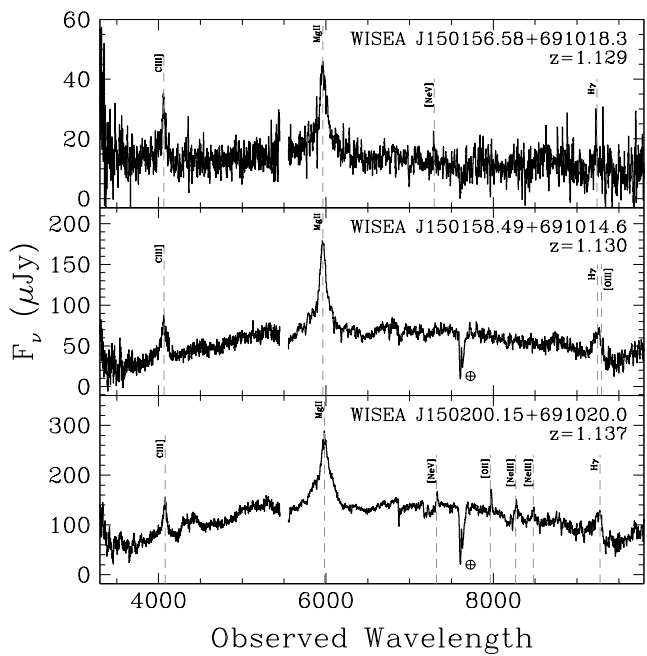}
    \caption{Optical spectra of the three WISE-selected AGN candidates
      associated with 2RXS~J150158.6+691029, from Palomar Observatory.
      All three are broad-lined quasars at similar redshifts, $z \sim
      1.13$. Note that we have not corrected the spectra for telluric
      absorption.}
    \label{fg:triplet_spectra}
  \end{center}
\end{figure}

We obtained optical spectroscopy of the three {\it WISE} AGN
candidates associated with one of these triplets,
2RXS~J150158.6+691029 (Fig. \ref{fg:triplet_image}), on UT 2016
October 2 with the optical dual-beam Double Spectrograph on the Hale
200-inch Telescope at Palomar Observatory, and we obtained optical
spectroscopy of the three WISE AGN candidates associated with another
of these triplets, 2RXS~J144427.2+311322, on UT 2017 April 28 with the
Low Resolution Imaging Spectrometer \citep[LRIS;][]{oke95} at Keck
Observatory.  The Palomar night had 1\arcsec\ seeing, with slight
cirrus in the morning, and we configured the instrument with the
1\farcs5 wide slit, the 5500~\AA\ dichroic, the 600 $\ell$ mm$^{-1}$
grating on the blue arm ($\lambda_{\rm blaze} = 4000$~\AA; spectral
resolving power $R \equiv \lambda / \Delta \lambda \sim 1200$), and
the 316 $\ell$ mm$^{-1}$ grating on the red arm ($\lambda_{\rm blaze}
= 7500$~\AA; $R \sim 1800$).  The Keck night was photometric with
sub-arcsecond seeing, and we configured the instrument with the
1\farcs0 wide slit, the 5600~\AA\ dichroic, the 600 $\ell$ mm$^{-1}$
grism on the blue arm ($\lambda_{\rm blaze} = 4000$~\AA; $R \sim
1200$), and the 400 $\ell$ mm$^{-1}$ grating on the red arm
($\lambda_{\rm blaze} = 8500$~\AA; $R \sim 1200$).  At Palomar, we
obtained two 900~s integrations, both at a position angle of
58.6$^\circ$.  The first integration simultaneously observed the two
AGN candidates to the East, while the second integration observed the
Western candidate.  At Keck, we obtained a single 300~s integration at
a position angle of 65.7$^\circ$, which simultaneously covered all
three WISE AGN candidates.  We processed all the data using standard
techniques within IRAF, and calibrated the spectra using standard
stars from \citet{Massey90} observed on the same nights. Table
\ref{tab:rosat_triplets} presents the results from the spectroscopy,
and Figure \ref{fg:triplet_spectra} presents the processed Palomar
spectroscopy, revealing three broad-lined quasars at $z \sim 1.13$;
the system observed at Keck turns out to be a quasar twin at $z \sim
1.75$ with a foreground interloper.  Higher resolution X-ray imaging
will be required to determine if the {\it ROSAT} emission detected in
these systems is due to associated hot intracluster media, is due to
emission from one (or more) of the quasars, or, least likely, is from
unrelated sources.

\section{Summary and Conclusions}

We have constructed large samples of robust AGN candidates across
$\sim$75\% of the sky selected solely from WISE observations. To
select the AGN candidates we followed the approach outlined by A13,
using only the W1 and W2 bands to maximize the number of
candidates. A13 proposed four selection criteria, two of which,
referred to as R90 and R75, optimized for reliability, producing AGN
samples that were respectively 90\% and 75\% reliable. The other two,
C90 and C75, were optimized for completeness, producing AGN samples
there were respectively 90\% and 75\% complete. These criteria were
calibrated for the data in the WISE All-Sky Data Release using a large
sample of AGN in the NDWFS Bo\"otes field. In this work we use the
more recent AllWISE data release, which is not only deeper in W1 and
W2 owing to the larger number of observations used, but also improved
several issues related to the data reduction in the initial data
release. Because of these differences, we recalibrated the selection
criteria following the same approach of A13 and found significant
differences for the optimal selection criteria. The reliability and
completeness optimized criteria are presented in equations
(\ref{eq:rel_sel}) and (\ref{eq:comp_sel}) respectively. We also
modified the functional form of the bright end of the reliability
optimized selection criteria, as the criteria provided by A13 was only
appropriate for the fainter end of the magnitude distribution.

We constructed two AGN catalogs, based on the R90 and C75 selection
criteria, respectively. In order to avoid contamination by artifacts
and by non-AGN sources in our Galaxy and in nearby galaxies, we
eliminated from the final catalogs all sources that i) are closer than
10 deg away from the GP and 30 deg from the GC, ii) are associated
with known PNe, iii) are associated with known H\,{\sc ii} and
star-forming regions, and iv) are associated to nearby galaxies. The
final R90 and C75 catalogs contain 4,543,530 and 20,907,127 AGN
candidates over an area of 30,093~deg$^2$, making them among the
largest quasar catalogs available. Through visual inspection of the
resulting catalog we find that a small number of spurious sources are
likely left due to artifacts related to the diffraction spikes of
bright stars, residuals left by the Solar System planets, and to
regions near the SAA with high Moon contamination. We then present a
few examples of scientific uses for these new AGN catalogs.

From the final R90 catalog we identify 687 AGN marked as highly
variable sources in the AllWISE catalog. Focusing on the 207 of these
sources that are within the FIRST survey footprint, we find that 162
(78\%) are detected by FIRST, indicating that they are likely
blazars. For the 45 remaining radio-undetected sources, we find that
32 have spectroscopic classifications in the literature, an additional
one has a photometric redshift, and we present spectroscopic
observations from the Palomar Observatory for two additional
sources. Of these 35, 28 are classified as AGN, four are classified as
galaxies and three are classified as stars. We find through visual
inspection of the spectra of the four sources classified as galaxies
that they likely host AGN.

Additionally, for one of these sources with archival SDSS spectra,
WISEA J142846.71+172353.1, we obtained a new epoch of spectroscopy. The
source faded by $\sim$1 magnitude between the WISE and NEOWISE-R
observations in both the W1 and W2 bands. The source is classified as
a QSO by SDSS, with a clear broad component to the H$\alpha$ emission
line, but only a narrow component to H$\beta$, implying an
intermediate-type classification. The SDSS observations were obtained
before the WISE data. In our new spectroscopic observations, obtained
after the WISE observations, we find that the broad component of
H$\alpha$ has disappeared, and the object is now consistent with a
type 2 classification.

We have carried out comparisons of our catalogs with other quasar
catalogs in the literature. We first compare our R90 catalog with the
also WISE-selected AGN catalog of \citet{secrest15}, which was built
using the \citet{mateos12} AGN selection criteria. Once we apply the
same spatial filtering done for our catalogs (see
\S\ref{ssec:spatial_filters}) and the same requirements discussed in
\S\ref{sec:agn_cats}, we find that only 3.7\% of the sources in their
catalog are not in the R90 sample. Conversely, the R90 sample is
approximately four times larger than the sample of
\citet{secrest15}. We also compare the R90 sample to the uXDQSOz
catalog of \citet{dipompeo15} and find that we primarily recover
sources with high $P_{\rm QSO}$, consistent with the high reliability
expected for the R90 sample. Additionally, we compare our AGN catalogs
with the SDSS DR12 quasar catalog and find that the R90 sample
recovers 30\% of the sources in the SDSS quasar catalog and 57\% of
the sources with WISE counterparts that meet our quality criteria. The
C75 sample, on the other hand, recovers 47\% of the SDSS quasar
sample, and 87\% of the sources with a WISE counterpart that meets our
quality criteria.

Finally, we compare our R90 sample with the 2RXS catalog, and find
that the R90 sample is approximately 250 times larger and recovers the
vast majority of the X-ray detected AGN in 2RXS. eROSITA, expected to
launch in the next year on the {\it Spectrum R\"ontgen Gamma}
satellite, will be approximately ten times more sensitive than {\it
  ROSAT} in the 0.5-2~keV band, and will provide the first sensitive,
all-sky survey in the 2-10~keV band \citep{merloni12}. \citet{stern12}
presents a detailed discussion of the expected comparison between
eROSITA- and WISE-identified AGN, where the latter is based on a
simpler, less comprehensive selection than presented here.  However,
the general result is expected to be the same, with WISE and eROSITA
both identifying most of the more luminous, less obscured quasars,
while WISE will do better at identifying obscured AGN and eROSITA will
do better at identifying lower-luminosity, unobscured AGN
\citep[see][for details]{stern12}.  Returning to the {\it ROSAT}
sample, we find that many of the 2RXS sources are matched to two or
more sources in the R90 catalog, which could be a promising sample for
identifying distant ($z>1$) X-ray emitting clusters. Applying a
conservative filtering to eliminate the possibility of artifacts
associated to the Moon contamination near the SAA latitudes (see
\S\ref{sssec:saa}) in these multiple matches, we find four 2RXS
sources each with three sources in the R90 sample within 30\arcsec. We
present spectra of two of these triple systems. One of these, 2RXS
J150158.6+691029, shows that all three sources are quasars at
$z\sim1.13$, suggestive of a group or proto-cluster at moderately high
redshifts.

We are currently conducting follow-up observations of a number of
interesting sources within the catalog. For example, \citet{assef16}
expands upon the analysis of radio-undetected AGN in the R90 sample
identified as highly variable in the AllWISE catalog. By 2019, we
anticipate a deeper all sky WISE-selected catalog than AllWISE will
become available, based on images combining data from WISE and
NEOWISE-R, from which larger AGN catalogs may be derived. We expect
the R90 and C75 WISE AGN catalogs will constitute a useful tool for
the astronomical community and be of use in a broad range of
applications.

\acknowledgments

We thank Mislav Balokovi\'c, Felix F\"urst, Brian Greffenstette,
Nikita Kamraj, George Lansbury, Sean Pike and Yanjun Xu for assisting
with the Palomar and Keck observations. We also thank Murray
Brightman, Matthew Graham, Christopher S. Kochanek and George
Djorgovski for helpful discussions that helped improve the paper. We
also thank the anonymous referee for useful comments and
suggestions. Some of the results in this paper have been derived using
the HEALPix \citep{gorski05} package. RJA was supported by FONDECYT
grant number 1151408. HJ is supported by Basic Science Research
Program through the National Research Foundation of Korea (NRF) funded
by the Ministry of Education (NRF-2017R1A6A3A04005158). This research
has made use of the NASA/IPAC Infrared Science Archive, which is
operated by the Jet Propulsion Laboratory, California Institute of
Technology, under contract with the National Aeronautics and Space
Administration. This publication makes use of data products from the
Wide-field Infrared Survey Explorer, which is a joint project of the
University of California, Los Angeles, and the Jet Propulsion
Laboratory/California Institute of Technology, and NEOWISE, which is a
project of the Jet Propulsion Laboratory/California Institute of
Technology. WISE and NEOWISE are funded by the National Aeronautics
and Space Administration.

\appendix
\section{AGN Catalog for Extended Sources}

The AGN catalogs constructed in \S\ref{sec:agn_cats} only consider
point sources from the AllWISE catalog. This is done because the
profile-fit photometry is optimized for point sources, and are
affected by different sets of systematics than for the aperture
photometry. As profile-fit measurements are the most robust for the
majority of the objects in AllWISE, it is reasonable for the
construction of the main catalogs to disregard extended
sources. However, AGN in WISE extended sources are interesting on
their own for a number of reasons. They are more likely to have higher
host contamination and hence are more likely to span lower Eddington
ratios. Furthermore dual AGN could potentially appear as extended WISE
sources if the separation of the nuclei is comparable to the PSF size.

In this appendix we provide catalogs of AGN identified in extended
sources using the same two selection criteria utilized for the point
source AGN catalogs, namely R90 and C75 (see
\S\ref{sec:agn_selection}). We construct these catalogs using almost
the same requirements on the WISE data as detailed in
\S\ref{sec:agn_cats}, except that we exchange the point source
requirement for a requirement that sources are flagged as extended
sources, and we do not filter out objects associated with 2MASS XSC
sources to not remove real, extended AGN from the sample. In other
words, we only consider sources with $S/N>5$ in W2, classified as
extended sources, not flagged as either artifacts or affected by
artifacts, and that pass all the spatial filters of
\S\ref{ssec:spatial_filters} except for the excising of 2MASS XSC
sources. The AllWISE Explanatory
Supplement\footnote{\url{http://wise2.ipac.caltech.edu/docs/release/allwise/expsup/sec2\_2.html\#psf\_fit}}
mentions that W1 images taken during the early part of the 3-Band Cryo
survey phase show a number of hard saturated pixels caused by the
rising temperatures. While the flux measurements are usually accurate,
the reduced $\chi^2$ of the profile fit in the W1 band can be
exceedingly large, leading to a spurious extended source
identification. Following the suggestion in the AllWISE Explanatory
Supplement, we eliminate all sources that have saturated pixels but a
W1 profile fitting magnitude fainter than the saturation limit (i.e.,
W1$>$8), as well as sources where the {\tt{ph\_qual}} flag for the W1
band is equal to {\tt{Z}}.

Despite the fact that the sources are extended to WISE, we apply the
R90 and C75 selection criteria using the profile-fitting
magnitudes. While the aperture magnitudes would provide a more
accurate measurement of the whole flux, the profile-fitting magnitudes
provide the better discrimination for AGN selection. In sources where
the W1 and W2 profiles are centrally concentrated, the profile-fitting
magnitudes provide a better representation of the central flux. In
sources where the extended regions are a more dominant fraction of the
total flux in these bands, the profile-fitting magnitudes will provide
a color more representative of the host galaxy, and hence these
sources will not meet our selection criteria, lowering the
completeness but without an effect on the reliability. However we
caution that the completeness and reliability estimates for our
criteria was done for point sources only, and hence may not be
accurate for these extended catalogs. Furthermore, we have not done as
careful a control for artifacts, which might be different for the
extended samples. 

We find that 20,645 and 26,331 extended sources are selected as AGN by
the R90 and C75 criteria, respectively. These sources are provided in
Tables \ref{tab:ext_r90_short} and \ref{tab:ext_c75_short}.

\input{tab1}
\pagebreak
\input{tab2}
\input{tab3}
\input{tab4}
\input{tab5}

\input{tab6}

\input{tab7}
\input{tab8}

\end{document}

%% file: tab1.tex
\begin{deluxetable*}{l c c c c c c c c c c c c}
  \tablecaption{R90 Catalog\label{tab:r90_short}}

  \tablehead{
    \colhead{WISE ID}&
    \colhead{RA} &
    \colhead{Dec} &
    \colhead{W1} &
    \colhead{$\sigma$(W1)} &
    \colhead{W2} &
    \colhead{$\sigma$(W2)} &
    \colhead{W3} &
    \colhead{$\sigma$(W3)} &
    \colhead{W4} &
    \colhead{$\sigma$(W4)} &
    \colhead{Moon-SAA} &
    \colhead{$\Sigma_{\rm pix}$}\\
    \colhead{(WISEA)} &
    \colhead{(deg)} &
    \colhead{(deg)} &
    \colhead{(mag)} &
    \colhead{(mag)} &
    \colhead{(mag)} &
    \colhead{(mag)} &
    \colhead{(mag)} &
    \colhead{(mag)} &
    \colhead{(mag)} &
    \colhead{(mag)} &
    \colhead{Flag} &
    \colhead{(deg$^{-2}$)}
  }

  \tabletypesize{\small}
  \tablewidth{0pt}
  \tablecolumns{13}

  \startdata
  J000000.00--165522.3   &   0.0000140 & --16.9228655   &   15.817 &   0.048 &   14.934 &   0.066 &   12.146 & \nodata &    8.851 & \nodata &   0 &   205\\
  J000000.01--422938.4   &   0.0000527 & --42.4940188   &   16.774 &   0.086 &   15.022 &   0.068 &   11.332 &   0.162 &    8.716 & \nodata &   0 &   176\\
  J000000.04+033452.5    &   0.0001897 & \phn\phn3.5812751 &   15.568 &   0.047 &   14.800 &   0.077 &   11.185 &   0.171 &    8.560 & \nodata &   0 &   200\\
  J000000.05--201340.3   &   0.0002095 & --20.2278802   &   17.705 &   0.225 &   16.055 &   0.187 &   12.034 & \nodata &    8.847 & \nodata &   0 &   310\\
  J000000.06--223834.6   &   0.0002592 & --22.6429645   &   16.429 &   0.078 &   15.079 &   0.086 &   11.943 &   0.341 &    8.610 & \nodata &   0 &   281\\
  J000000.06--473835.1   &   0.0002617 & --47.6430989   &   14.086 &   0.027 &   13.233 &   0.028 &\phn9.987 &   0.048 &    7.551 &   0.155 &   0 &   157\\
  J000000.08+165703.8    &   0.0003373 & \phn16.9510671 &   17.022 &   0.127 &   15.692 &   0.134 &   11.894 &   0.333 &    8.309 & \nodata &   0 &   210\\
  J000000.09--293647.0   &   0.0003889 & --29.6130691   &   16.248 &   0.064 &   15.226 &   0.083 &   12.108 &   0.276 &    8.700 & \nodata &   0 &   200\\
  J000000.12--324059.2   &   0.0005112 & --32.6831183   &   17.134 &   0.125 &   15.862 &   0.132 &   12.411 & \nodata &    9.189 & \nodata &   0 &   281\\
  J000000.14+190345.9    &   0.0006023 & \phn19.0627595 &   18.092 & \nodata &   16.126 &   0.191 &   12.571 & \nodata &    9.054 & \nodata &   0 &   172\\
  \enddata

  \tablecomments{The magnitudes and errors shown correspond to the
    profile-fitting measurements in the AllWISE catalog. Undetected
    sources in a given band lack a magnitude uncertainty measurement
    and the magnitude column shows a 95\% confidence upper bound. The
    quantity $\Sigma_{\rm pix}$ is defined
    in \S\ref{sssec:planets}. (This table is available in its entirety
    in a machine-readable form in the online journal. A portion is
    shown here for guidance regarding its form and content.)}

\end{deluxetable*}

%% file: tab2.tex
\begin{deluxetable*}{l c c c c c c c c c c c c}

  \tablecaption{C75 Catalog\label{tab:c75_short}}

  \tablehead{
    \colhead{WISE ID}&
    \colhead{RA} &
    \colhead{Dec} &
    \colhead{W1} &
    \colhead{$\sigma$(W1)} &
    \colhead{W2} &
    \colhead{$\sigma$(W2)} &
    \colhead{W3} &
    \colhead{$\sigma$(W3)} &
    \colhead{W4} &
    \colhead{$\sigma$(W4)} &
    \colhead{Moon-SAA} &
    \colhead{$\Sigma_{\rm pix}$}\\
    \colhead{(WISEA)} &
    \colhead{(deg)} &
    \colhead{(deg)} &
    \colhead{(mag)} &
    \colhead{(mag)} &
    \colhead{(mag)} &
    \colhead{(mag)} &
    \colhead{(mag)} &
    \colhead{(mag)} &
    \colhead{(mag)} &
    \colhead{(mag)} &
    \colhead{Flag} &
    \colhead{(deg$^{-2}$)}
  }

  \tabletypesize{\small}
  \tablewidth{0pt}
  \tablecolumns{13}

  \startdata
  J000000.00--314627.5   &   0.0000000 & --31.7743100   &   16.874 &   0.099 &   16.093 &   0.168 &   12.211 & \nodata &    9.099 & \nodata &   0 &   896\\
  J000000.00--485007.6   &   0.0000076 & --48.8354646   &   17.411 &   0.144 &   16.350 &   0.206 &   11.834 & \nodata &    8.542 & \nodata &   0 &   920\\
  J000000.00--165522.3   &   0.0000140 & --16.9228655   &   15.817 &   0.048 &   14.934 &   0.066 &   12.146 & \nodata &    8.851 & \nodata &   0 &   729\\
  J000000.01--422938.4   &   0.0000527 & --42.4940188   &   16.774 &   0.086 &   15.022 &   0.068 &   11.332 &   0.162 &    8.716 & \nodata &   0 &   896\\
  J000000.01--323326.5   &   0.0000589 & --32.5573670   &   16.771 &   0.091 &   16.002 &   0.166 &   12.545 & \nodata &    9.133 &   0.468 &   0 &   963\\
  J000000.01--112405.6   &   0.0000639 & --11.4015677   &   17.179 &   0.140 &   16.153 &   0.199 &   12.517 & \nodata &    9.035 & \nodata &   0 &   620\\
  J000000.01+350440.6    &   0.0000738 & \phn35.0779461 &   16.990 &   0.109 &   16.035 &   0.170 &   12.066 & \nodata &    9.083 & \nodata &   0 &   477\\
  J000000.02--485353.6   &   0.0001188 & --48.8982362   &   16.537 &   0.073 &   15.815 &   0.138 &   12.146 & \nodata &    8.554 & \nodata &   0 &   920\\
  J000000.03+140926.9    &   0.0001278 & \phn14.1574789 &   17.271 &   0.149 &   16.094 &   0.194 &   12.386 & \nodata &    8.188 & \nodata &   0 &   596\\
  J000000.03--191610.5   &   0.0001360 & --19.2696075   &   16.617 &   0.091 &   15.652 &   0.140 &   12.459 & \nodata &    8.962 & \nodata &   0 &   581\\
  \enddata

  \tablecomments{The magnitudes and errors shown correspond to the
    profile-fitting measurements in the AllWISE catalog. Undetected
    sources in a given band lack a magnitude uncertainty measurement
    and the magnitude column shows a 95\% confidence upper bound. The
    quantity $\Sigma_{\rm pix}$ is defined
    in \S\ref{sssec:planets}. (This table is available in its entirety
    in a machine-readable form in the online journal. A portion is
    shown here for guidance regarding its form and content.)}

\end{deluxetable*}

%% file: tab3.tex
\begin{deluxetable}{l c c c c}

  \tablecaption{HEALPix Pixel Surface Density $\Sigma_{\rm pix}$
    (deg$^{-2}$)\label{tab:healpix_dens}}

  \tablehead{
    \colhead{Sample}&
    \multicolumn{4}{c}{Percentiles}\\
    \colhead{}&
    \colhead{Median}&
    \colhead{90$^{\rm th}$}&
    \colhead{95$^{\rm th}$}&
    \colhead{99$^{\rm th}$}
  }

  \tabletypesize{\small}
  \tablewidth{0pt}
  \tablecolumns{5}

  \startdata
  R90, W2 S/N$>$5  &    143  & \phn219 & \phn248  & \phn319 \\
  R90, W2 S/N$>$10 & \phn81  & \phn124 & \phn133  & \phn157 \\
  C75, W2 S/N$>$5  &    643  &    1096 &    1249  &    1516 \\
  C75, W2 S/N$>$10 &    119  & \phn238 & \phn305  & \phn448 \\
  \enddata


\end{deluxetable}

%% file: tab4.tex
\begin{deluxetable}{l c l l}

  \tablecaption{Spectroscopic Properties of Radio-Quiet, Highly
    Variable WISE AGN\label{tab:var_rq_specs}}

  \tablehead{
    \colhead{WISE ID}&
    \colhead{Redshift}&
    \colhead{Classification}&
    \colhead{Ref}\\
    \colhead{(WISEA)}&
    \colhead{}&
    \colhead{}&
    \colhead{}
  }

  \tabletypesize{\small}
  \tablewidth{0pt}
  \tablecolumns{4}

  \startdata
  J000011.72+052317.4  & 0.0400 & Seyfert 1    & SIMBAD\\
  J014004.69--094230.4 & 0.1461 & QSO          & SDSS\\
  J090546.35+202438.2  &\nodata & Carbon Star  & SIMBAD\\
  J091225.00+061014.8  & 0.1453 & Galaxy\tablenotemark{$\dagger$} & SDSS\\
  J094806.56+031801.7  & 0.2073 & QSO          & SDSS\\
  J095824.97+103402.4  & 0.0417 & Galaxy AGN   & SDSS\\
  J100933.13+232255.7  & 0.0719 & Galaxy AGN   & SDSS\\
  J104241.08+520012.8  & 0.1365 & QSO          & SDSS\\
  J112537.83+212042.2  & 0.0894 & QSO          & SDSS\\
  J130155.84+083631.7  &\nodata & Carbon Star  & SIMBAD\\
  J130716.98+450645.3  & 0.0843 & QSO          & SDSS\\
  J130819.11+434525.6  & 0.0365 & Galaxy AGN   & SDSS\\
  J140033.66+154432.1  & 0.2152 & QSO          & SDSS\\
  J141053.43+091027.0  & 0.1781 & QSO          & SDSS\\
  J141105.45+294211.8  & 0.0724 & QSO          & SDSS\\
  J141758.60+091609.7  & 0.1389 & QSO          & SDSS\\
  J142747.45+165206.0  & 0.1435 & QSO          & SDSS\\
  J142846.71+172353.1  & 0.1040 & QSO          & SDSS\\
  J144039.30+612748.1  & 0.0811 & QSO          & SDSS\\
  J144131.81+321612.9  & 0.1993 & QSO          & SDSS\\
  J144439.59+351304.7  & 0.0790 & Galaxy\tablenotemark{$\dagger$}  & SDSS\\
  J144510.14+304957.1  & 0.2754 & QSO          & SDSS\\
  J144603.98--013203.4 & 0.0840 & Galaxy AGN   & SDSS\\
  J145222.03+255152.0  & 0.1204 & QSO          & SDSS\\
  J150954.94+203619.6  & 0.4149\tablenotemark{$\ddagger$}  & Possible AGN & SIMBAD\\
  J151215.73+020316.9  & 0.2199 & Galaxy AGN   & SDSS\\
  J151444.52+364237.9  & 0.1944 & QSO          & SDSS\\
  J151518.56+312937.5  & 0.1036 & QSO          & SDSS\\
  J155223.29+323455.0  & 0.1277 & Galaxy\tablenotemark{$\dagger$} & SDSS\\
  J161846.36+510035.1  & 0.3189 & QSO          & SDSS\\
  J162140.25+390105.1  & 0.0642 & Galaxy AGN   & SDSS\\
  J163518.38+580854.6  &\nodata & Star         & SIMBAD\\
  J213604.22--050152.0 & 0.1284 & Galaxy\tablenotemark{$\dagger$} & SIMBAD\\
  \enddata

  \tablenotetext{$\dagger$}{Although the object is classified as a
    galaxy in SDSS or SIMBAD, the H$\alpha$ emission line shows a
    broad base suggesting the presence of an AGN.}
  \tablenotetext{$\ddagger$}{Photometric redshift. No spectroscopic
    classification is available for this object.}

\end{deluxetable}

%% file: tab5.tex
\begin{deluxetable}{l c l}

  \tablecaption{Spectroscopic Follow-up of Highly Variable WISE
    AGN\label{tab:var_rq_specs_palomar}}

  \tablehead{
    \colhead{WISE ID}&
    \colhead{Redshift}&
    \colhead{Classification}\\
    \colhead{(WISEA)}&
    \colhead{}&
    \colhead{}
  }

  \tabletypesize{\small}
  \tablewidth{0pt}
  \tablecolumns{3}

  \startdata
   {\it{Undetected by FIRST}}\\
   J015858.48+011507.6  & 0.184 & Type 2 AGN\\
   J101536.17+221048.9  & 0.555 & Red Type 1 AGN\\
   \\
   {\it{Detected by FIRST}}\\
   J090931.55--011233.3 & 0.201 & \\
   J095528.76+572837.2  & \nodata & Blazar?\\
   \\
   {\it{Outside FIRST}}\\
   J051939.78+160044.0  & \nodata & Galactic CV\\
   \enddata

\end{deluxetable}

%% file: tab6.tex
\begin{deluxetable}{llccc}

  \tablewidth{0pt}

  \tablecaption{Quasar triplets with {\it ROSAT} counterparts\label{tab:rosat_triplets}}

  \tablehead{
    \colhead{2RXS source} &
    \colhead{WISE source} &
    \colhead{W1} &
    \colhead{W2} &
    \colhead{$z$}}
  \startdata
  2RXS J144427.2+311322 & WISEA J144426.93+311307.8 & 14.87 & 13.89 & 0.449\\
                        & WISEA J144427.97+311313.9 & 15.91 & 14.68 & 1.798\\
                        & WISEA J144429.40+311321.2 & 13.98 & 12.62 & 1.730\\
  \\
  2RXS J150158.6+691029 & WISEA J150156.58+691018.3 & 15.78 & 14.78 & 1.129\\
                        & WISEA J150158.49+691014.6 & 14.68 & 13.54 & 1.130\\
                        & WISEA J150200.15+691020.0 & 14.93 & 13.58 & 1.137\\
  \\
  2RXS J162949.1+052341 & WISEA J162948.86+052353.1 & 15.10 & 14.41 & \\
                        & WISEA J162949.68+052358.0 & 13.94 & 12.89 & \\
                        & WISEA J162951.29+052328.1 & 15.92 & 14.88 & \\
  \\
  2RXS J220938.7-332250 & WISEA J220936.03-332247.2 & 15.97 & 14.86 & \\
                        & WISEA J220938.34-332237.6 & 13.43 & 12.17 & \\
                        & WISEA J220940.80-332244.4 & 16.60 & 15.01 & \\
  \enddata

  \tablecomments{The redshift of the brightest 14-hr source is from
    SDSS, while redshifts of the other two are from our Keck
    follow-up. The three redshifts for the 15-hr sources are from our
    Palomar follow-up.}

\end{deluxetable}

%% file: tab7.tex
\begin{deluxetable}{l c c c c c c c c c c c}

  \tablecaption{R90 Catalog of Extended Sources\label{tab:ext_r90_short}}
  
  \tablehead{
    \colhead{WISE ID}&
    \colhead{RA} &
    \colhead{Dec} &
    \colhead{W1} &
    \colhead{$\sigma$(W1)} &
    \colhead{W2} &
    \colhead{$\sigma$(W2)} &
    \colhead{W3} &
    \colhead{$\sigma$(W3)} &
    \colhead{W4} &
    \colhead{$\sigma$(W4)} &
    \colhead{Moon-SAA Flag}\\
    \colhead{(WISEA)} &
    \colhead{(deg)} &
    \colhead{(deg)} &
    \colhead{(mag)} &
    \colhead{(mag)} &
    \colhead{(mag)} &
    \colhead{(mag)} &
    \colhead{(mag)} &
    \colhead{(mag)} &
    \colhead{(mag)} &
    \colhead{(mag)} &
    \colhead{}
  }

  \tabletypesize{\small}
  \tablewidth{0pt}
  \tablecolumns{12}

  \startdata
  J000005.02+085706.6    &   0.0209340 &\phn\phn8.9518348 &   14.745 &   0.032 &   13.894 &   0.040 &   11.406 &   0.193 &    8.506 &   0.370 &   0\\
  J000009.53--455127.8   &   0.0397123 &     --45.8577390 &   14.312 &   0.027 &   13.338 &   0.030 &   10.573 &   0.095 &    7.968 &   0.217 &   0\\
  J000011.06+052307.7    &   0.0460950 &\phn\phn5.3854886 &   14.443 &   0.029 &   13.743 &   0.038 &   10.588 &   0.097 &    7.254 &   0.125 &   0\\
  J000020.23+221358.5    &   0.0843316 &   \phn22.2329366 &   18.524 & \nodata &   15.972 &   0.190 &   12.154 & \nodata &    8.732 & \nodata &   0\\
  J000021.94+323159.5    &   0.0914206 &   \phn32.5332050 &   16.325 &   0.066 &   14.996 &   0.074 &   11.879 &   0.237 &    9.035 &   0.534 &   0\\
  J000024.71--523254.8   &   0.1029903 &     --52.5485593 &   15.092 &   0.101 &   14.357 &   0.095 &   11.591 & \nodata &    8.573 & \nodata &   0\\
  J000027.12+050511.5    &   0.1130353 &\phn\phn5.0865358 &   15.410 &   0.123 &   14.245 &   0.108 &   11.339 & \nodata &    8.714 & \nodata &   0\\
  J000045.12--380735.3   &   0.1880166 &     --38.1264753 &   15.568 &   0.040 &   14.272 &   0.044 &   10.703 &   0.087 &    8.725 &   0.305 &   0\\
  J000051.46--380145.8   &   0.2144444 &     --38.0293889 &   14.986 &   0.034 &   13.734 &   0.032 &   10.158 &   0.054 &    7.644 &   0.138 &   0\\
  J000058.83--245451.2   &   0.2451389 &     --24.9142450 &   14.923 &   0.035 &   14.217 &   0.047 &   11.754 &   0.222 &    8.811 &   0.442 &   0\\
  \enddata

  \tablecomments{The magnitudes and errors shown correspond to the
    profile-fitting measurements in the AllWISE catalog. Undetected
    sources in a given band lack a magnitude uncertainty measurement
    and the magnitude column shows a 95\% confidence upper bound. (This
    table is available in its entirety in a machine-readable form in
    the online journal. A portion is shown here for guidance regarding
    its form and content.)}

\end{deluxetable}

%% file: tab8.tex
\begin{deluxetable}{l c c c c c c c c c c c}

  \tablecaption{C75 Catalog of Extended Sources\label{tab:ext_c75_short}}
  
  \tablehead{
    \colhead{WISE ID}&
    \colhead{RA} &
    \colhead{Dec} &
    \colhead{W1} &
    \colhead{$\sigma$(W1)} &
    \colhead{W2} &
    \colhead{$\sigma$(W2)} &
    \colhead{W3} &
    \colhead{$\sigma$(W3)} &
    \colhead{W4} &
    \colhead{$\sigma$(W4)} &
    \colhead{Moon-SAA Flag}\\
    \colhead{(WISEA)} &
    \colhead{(deg)} &
    \colhead{(deg)} &
    \colhead{(mag)} &
    \colhead{(mag)} &
    \colhead{(mag)} &
    \colhead{(mag)} &
    \colhead{(mag)} &
    \colhead{(mag)} &
    \colhead{(mag)} &
    \colhead{(mag)} &
    \colhead{}
  }

  \tabletypesize{\small}
  \tablewidth{0pt}
  \tablecolumns{12}

  \startdata
  J000005.02+085706.6    &   0.0209340 &\phn\phn8.9518348 &   14.745 &   0.032 &   13.894 &   0.040 &   11.406 &   0.193 &    8.506 &   0.370 &   0\\
  J000009.53--455127.8   &   0.0397123 &     --45.8577390 &   14.312 &   0.027 &   13.338 &   0.030 &   10.573 &   0.095 &    7.968 &   0.217 &   0\\
  J000020.23+221358.5    &   0.0843316 &   \phn22.2329366 &   18.524 & \nodata &   15.972 &   0.190 &   12.154 & \nodata &    8.732 & \nodata &   0\\
  J000021.94+323159.5    &   0.0914206 &   \phn32.5332050 &   16.325 &   0.066 &   14.996 &   0.074 &   11.879 &   0.237 &    9.035 &   0.534 &   0\\
  J000024.71--523254.8   &   0.1029903 &     --52.5485593 &   15.092 &   0.101 &   14.357 &   0.095 &   11.591 & \nodata &    8.573 & \nodata &   0\\
  J000025.30--400339.3   &   0.1054477 &     --40.0609388 &   16.724 &   0.089 &   15.766 &   0.136 &\phn9.776 &   0.067 &    6.116 &   0.063 &   0\\
  J000027.12+050511.5    &   0.1130353 &\phn\phn5.0865358 &   15.410 &   0.123 &   14.245 &   0.108 &   11.339 & \nodata &    8.714 & \nodata &   0\\
  J000045.12--380735.3   &   0.1880166 &     --38.1264753 &   15.568 &   0.040 &   14.272 &   0.044 &   10.703 &   0.087 &    8.725 &   0.305 &   0\\
  J000047.52+274212.5    &   0.1980083 &   \phn27.7034994 &   16.254 &   0.056 &   15.498 &   0.114 &   12.521 & \nodata &    8.671 & \nodata &   0\\
  J000051.46--380145.8   &   0.2144444 &     --38.0293889 &   14.986 &   0.034 &   13.734 &   0.032 &   10.158 &   0.054 &    7.644 &   0.138 &   0\\
  \enddata

  \tablecomments{The magnitudes and errors shown correspond to the
    profile-fitting measurements in the AllWISE catalog. Undetected
    sources in a given band lack a magnitude uncertainty measurement
    and the magnitude column shows a 95\% confidence upper
    bound. (This table is available in its entirety in a
    machine-readable form in the online journal. A portion is shown
    here for guidance regarding its form and content.)}

\end{deluxetable}